%% file: main.tex
\documentclass[letterpaper,twocolumn,10pt]{article}
\usepackage{usenix2019_v3}

\usepackage{tikz}

\usepackage{textcomp} 
\usepackage[english]{babel}
\usepackage{times}
\usepackage{microtype}
\usepackage{amsfonts,amsmath, amssymb}
\usepackage{balance}
\usepackage{graphicx}
\usepackage{multirow}
\usepackage{paralist}
\usepackage[font=small,labelfont=bf]{caption}
\usepackage{subcaption}
\usepackage{xspace}
\usepackage{booktabs}
\usepackage{listings}
\usepackage{hyperref}
\usepackage{listings}
\usepackage{enumitem}
\usepackage{xcolor}
\usepackage{siunitx}
\usepackage{algorithm}
\floatname{algorithm}{Component}
\usepackage[noend]{algpseudocode}
\usepackage{array}
\usepackage{balance}
\usepackage[breakable, theorems, skins]{tcolorbox}
\usepackage{float}

\usepackage{csquotes}
\newcommand{\name}{\textsc{RouteScout}\xspace}
\newcommand{\solver}{\textit{Solver}\xspace}

\newcommand{\Solver}{\textit{Solver}\xspace}

\newcommand{\fwd}{\textit{forwarding Selector}\xspace}
\newcommand{\mnt}{\textit{monitoring Selector}\xspace}

\newcommand{\acc}{\textit{Accumulator}\xspace}
\newcommand{\selector}{\textit{Selector}\xspace}

\newcommand{\coun}{\textit{Counter}\xspace}
\newcommand{\mon}{\textit{monitor}\xspace}

\newcommand{\agg}{\textit{aggregator}\xspace}

\newcommand{\delm}{\textit{Delay monitor}\xspace}
\newcommand{\losm}{\textit{Loss monitor}\xspace}
\newcommand{\dela}{\textit{Delay aggregator}\xspace}

\newcommand{\ed}[1]{\textsf{\textcolor{red}{[#1]}}}
\newcommand*\todo[1]{\textcolor{red}{\textbf{TODO:} #1}}%

\newcommand{\laurent}[1]{\ed{#1---\textbf{LV}}}

\newcommand{\remove}[1]{}
\newcommand{\myitem}[1]{\vspace*{0.02in}\noindent\textbf{#1}}

\usepackage{enumitem}
\setlist[itemize]{topsep=1pt}
\setlist[enumerate]{topsep=1pt}
\setlist[itemize]{itemsep=0pt}
\setlist[enumerate]{itemsep=0pt}

\makeatletter
\newcommand{\labeledItem}[1][]{
        \protected@edef\@currentlabel{#1}%
\item[#1]
}
  \makeatother

\newcommand*\xor{\mathbin{\oplus}}

\newcommand{\ie}{\emph{i.e.,~}}
\newcommand{\eg}{\emph{e.g.,~}}
\definecolor{darkgreen}{rgb}{0,0.5,0}
\definecolor{brown}{rgb}{0.7,0.3,0}
\definecolor{darkblue}{rgb}{0,0,0.5}
\definecolor{darkred}{rgb}{0.5,0,0}
\definecolor{mygray}{gray}{0.5}
\newcommand{\parab}[1]{\vspace{0.03in}\noindent\textbf{#1}}

\begin{document}

\date{}


\title{Performance-Driven Internet Path Selection}

\author{
{\rm Maria Apostolaki}\\
ETH Z{\"u}rich
\and
{\rm Ankit Singla }\\
 ETH Z{\"u}rich
\and
{\rm Laurent Vanbever}\\
ETH Z{\"u}rich
} 

\maketitle

\input{abstract}

\input{introduction}
\input{motivation}

\input{overview}
\input{data}
\input{implementation}

\input{multi_control}

\input{dataEval}

\input{control_eval}

\input{case_study}

\input{conclusion}

\section*{Acknowledgments}

This work was supported by a Swiss National Science Foundation Grant
(``Data-Driven Internet Routing'', \#200021-175525).

\bibliographystyle{plain}
\bibliography{main}


\end{document}

%% file: abstract.tex
\begin{abstract}
Internet routing can often be sub-optimal, with the chosen routes providing
worse performance than other available policy-compliant
routes. This stems from the lack of visibility into
route performance at the network layer. While this is an old
problem, we argue that recent advances in programmable hardware finally open up
the possibility of performance-aware routing in a deployable, BGP-compatible
manner.

We introduce \name, a hybrid hardware/software system supporting
performance-based routing at ISP scale. In the data plane, \name
leverages P4-enabled hardware to monitor performance across policy-compliant route
choices for each destination, at line-rate and with a small memory footprint.
\textsc{RouteScout}'s control plane then asynchronously pulls aggregated performance metrics to synthesize a performance-aware forwarding policy. 

We show that \name can monitor performance across most of an ISP's traffic, using only 4 MB of memory. Further, its control can flexibly satisfy a variety of operator objectives, with sub-second operating times.
\end{abstract}

%% file: introduction.tex
\section{Introduction}
\label{sec:intro}





Internet routing uses cost-driven policies to select \emph{one} interdomain
path per destination along which to direct traffic. To select one path amongst
multiple policy-compliant ones, the Internet's Border Gateway Protocol (BGP)
uses particularly crude criteria rather than dynamically optimizing for performance. For instance, BGP will favor paths crossing fewer networks or paths crossing networks whose identifiers are smaller.\footnote{One of BGP tie-breaking criteria is indeed to prefer routes announced by the router with the smallest IP address~\cite{rfc4271}.} As a result, BGP selects routes that are often sub\-optimal in terms of throughput, latency, and reliability.

This problem is far from new and the sub-optimality of Internet routing is
long-established~\cite{savage1999detour,spring2003causes,
tangmunarunkit2001internet}. Yet, despite several strong
attempts~\cite{savage1999detour, andersen2002resilient, akella2003measurement,
akella2004multihoming, akella2008performance, valancius2013quantifying}, few
practical progresses have been made. The problem is that enabling performance-aware routing is particularly challenging requiring: scalable monitoring of path performance, handling path dynamics, stability and correctness of routing, and insurmountable resistance to any approach incompatible with BGP.


Despite the problem's difficulty and its long history, we posit its time
to revisit this problem, for three reasons. 

First, Internet application requirements have evolved, with a sharper focus on \emph{reliably high} network performance.
For hyperscale Web services with numerous well-connected points-of-presence across the globe, BGP is, in fact, good enough
\emph{most of the time}~\cite{arnold2019beating}. 
However, even in these best-case environments, the benefits of reducing tail latency and performance variability in response to transient congestion
are valuable enough for providers like Google and Facebook to invest in performance-aware routing~\cite{googEspresso,fbEdgeConnect}.
Google's Espresso showed that being able to dynamically reroute around transient congestion
improved mean time between rebuffers in their video service by $35$--$170\%$~\cite{googEspresso}. Espresso explicitly pins these gains on
being able to dynamically respond to performance \emph{variability} across paths (rather than just average-case improvement from an one-time evaluation), thus underscoring the need for making path decisions based on \emph{continuous assessments of the changing performance} of paths. 
Beyond Web services, other applications are even more demanding: in gaming, even small latency overheads can put players at a disadvantage~\cite{riotGames}. 
The importance of tail latency as opposed to mean latency is also demonstrated on CDN's efforts to improve latency of the worst-performing clients\cite{chen2015end}.
Thus, \emph{if} performance-aware routing were practical, the benefits would justify significant design effort.

Second, the available paths are increasingly diverse, due to increased peering and the establishment of Internet Exchange Points (IXPs), which simply did not exist at the time of BGP's first design iteration (1989). 
Further, if plans for satellite-based global Internet connectivity~\cite{starlink, amazon_news} come to fruition, the performance gap across different paths will also increase. Two teams of researchers have separately argued in recent position papers~\cite{debopam_hotnets2018,tobias_hotnets2018} that these satellite systems exhibit continuous changes in both the performance and availability of routes, and thus, will pose challenges to the performance-oblivious and slow-to-converge BGP routing.

Third, the recent development of programmable switches that allow line-rate, per-packet data plane operations enables new design primitives. These heretofore unavailable primitives, as we shall show, drastically improve our ability to both evaluate and control multiple candidate routes.

Motivated by the above factors, we present \textbf{\name}, a novel software-hardware co-design for 
performance-aware routing. 
\name's data plane estimates loss and delay along different policy-compliant next-hop
routes for different destinations. It leverages probabilistic data structures in programmable switches
to aggregate delay and loss measurements on a per destination-nexthop granularity. This
in-data-plane aggregation eliminates the necessity of mirroring traffic
 to more powerful general purpose hardware, thus alleviating:
(a) bandwidth and compute overheads; and (b) deterioration in monitoring
capabilities when they are most needed, under congestion.
Past methods (\S\ref{sec:rs_need}) are incapable of producing
such accurate, high-coverage, real-time, and low-overhead performance measurements
for multiple candidate nexthops for a large number of destinations. 

The succinct measurements allow \name's control plane to evaluate
multiple policy-compliant candidate paths by measuring their
performance systematically for small slices of live traffic. 
\name then encodes the best path choices in the data plane using 
a small memory footprint. \name enforces those choices gradually, 
while continually monitoring performance to avoid self-induced congestion and, therefore, oscillations~\cite{gao2006avoiding}.


While \name could be used by any Autonomous System (AS), for tractability of control, we trim the problem's scope: we take
the perspective of a stub AS which offers no transit services to other ASes. This eliminates the risk of conflicting decision-making leading to transient loops and instability. We humbly suggest that this ``relaxation'' still leads to a highly non-trivial and useful setting: stubs comprise $85\%$  of all ASes;\footnote{A likely low estimate, computed from CAIDA's AS-level topology~\cite{caida_as_level}.} the majority of stubs are multi-homed and virtually all Internet traffic and originates from some stub. In addition, despite sitting at the edge of the Internet, stubs often know several paths to reach each destination: our measurements on CAIDA AS-level topologies~\cite{caida_as_level} reveal that the majority of them (55\%) can use at least two equally-preferred paths for at least 80\% of the destinations.\footnote{For each stub we calculated the number of BGP-equivalent paths for $1000$ randomly selected destination prefixes, following~\cite{GaoRexford_StableInternetRouting_2001}.} Stubs also tend to connect with their neighbors via redundant links, further increasing path diversity~\cite{10.1145/1644893.1644937}.
Finally, while \name can only control paths \emph{from} the stub, not towards it, the resulting reductions in round-trip time, and being able to avoid congestion/failures at least in one direction, are still valuable improvements. 


\name is carefully designed to run on
available programmable switches, respecting constraints on memory,
operations per packet, memory accesses per packet, and constraints on
accesses to memory blocks across pipeline stages. It requires no coordination
across ASes and works over unmodified BGP. Within an AS, it yields benefits starting with only
one programmable switch deployed at the edge.

\vspace{0.08in}
\noindent Our main \textbf{contributions} are the following:
\vspace{-0.1in}

\begin{itemize}[leftmargin=*]
\setlength\itemsep{0pt}
\item \name, a system capable of rerouting traffic to test the performance of alternative routes to each destination prefix, in a controlled and automated manner.

\item Methods to compute delay and loss rates across different paths that are accurate and effective, while respecting the constraints of data-plane hardware.

\item Efficient interconnection between the control and
data plane that allows: (a) fast, fine-grained, and asynchronous changes in the forwarding and monitoring policy;
(b) fast, fine-grained, and low-bandwidth retrieval of statistics.

\item An implementation of \name on a Barefoot Tofino switch~\cite{tofino}, with an evaluation of its control- and data-plane. 

\end{itemize}

%% file: motivation.tex
\section{Motivation}
\label{sec:rs_need}
Performance-aware routing is an old problem~\cite{akella2004multihoming,akella2008performance,savage1999detour,spring2003causes, tangmunarunkit2001internet}, with several known solutions of varying ambition and complexity. Early work~\cite{goldenberg2004optimizing} narrowly targeted multi-homed end-users with perfect visibility over their performance,  cost being their first priority, and direct links the only possible bottleneck. TeXCP~\cite{kandula2005walking} and MATE~\cite{elwalid2001mate} focused on intra-domain routing, splitting traffic across already setup tunnels. We would instead like to tackle the problem from the perspective of an AS picking routes to external destinations, with no end-host control, and only observing its own traffic. 
In this setting, we discuss several alternatives for monitoring path performance, whose limitations make the case for \name.

\myitem{Active probing:} One can actively probe routes~\cite{cisco-pfr, ip-sla}. While this approach can be effective in the intra-domain setting, where recent work~\cite{hsu2019contra, INT} used specially crafted probes to monitor performance, it is insufficient for our inter-domain context as probes may not be representative of real traffic's performance --- the volume of probing traffic is likely orders of magnitude less than the actual traffic, and some ISPs are known to treat probing traffic preferentially~\cite{clark_congestion_interdomain}. Several systems propose to address some of these issues by collecting and combining measurements from end-users~\cite{netprofiler, spand}. However, requiring large numbers of cooperative users makes bootstrapping hard.



\myitem{Passive sampling:} Gathering statistics on live traffic is possible using sampling with sFlow~\cite{sflow} or NetFlow~\cite{netflow}. However, sampling simply does not capture performance --- measuring these metrics requires capturing state across particular packets per flow (\S\ref{subsec:delay}, \S\ref{ssec:loss_component}), not arbitrary random samples.


\myitem{Mirroring:} While mirroring obviously captures the requisite information, it does not scale and is inflexible. 
To avoid congestion from mirrored traffic, one can rate-limit it, but this has limitations similar to sampling: naive rate-limiting will discard arbitrary packets across flows, impairing loss and delay estimation. Alternatively, one can target mirroring more narrowly, with systems like Everflow~\cite{everflow} and Stroboscope~\cite{stroboscope}. However, for continuous, high-coverage monitoring across Internet prefixes and potential next-hops, such methods would require a large and constantly changing set of monitoring rules in network devices. Further, even if we could dynamically match on a given number of flows per prefix and mirror only those (\eg with programmable switches to store flow identifiers), the mirrored traffic will still be burdensome. 

As an illustration, consider an operator who wants to monitor the performance for traffic sent to $1000$ destinations over only $2$ alternative next-hops and by mirroring only $50$ flows per destination-next-hop pair. At the mean flow rate observed in CAIDA traces~\cite{caida}, we find that such a design would require mirroring $25.7$~Gbps of traffic. In contrast, by aggregating measurements directly in the data plane, \name generates $108.4$~kbps in performance reports, \ie at $287$,$000$$\times$ higher efficiency.


\paragraph{End-system monitoring:} Google~\cite{googEspresso} and Facebook~\cite{fbEdgeConnect} have recently shared their solutions for path-aware routing. These approaches leverage their unique control: one end of the monitored connections terminates at their own powerful servers, and the other at a client application that also supplies performance data. This is obviously infeasible for ASes.


\myitem{Sketches:} Sketches~\cite{flowradar,DREAM,UnivMon,TwoLevel,yang2018elastic,kilpi2008ip} offer aggregate estimates for packet/flow counts and size distributions, but do not capture latency and loss across routes.

\vspace{0.1in}
\name exploits \textbf{programmable switches} that open up avenues unavailable to past efforts. To the best of our knowledge, no prior work leveraging programmable switches fully addresses either the sensing / monitoring or the control necessary for performance-aware routing.
Blink~\cite{holterbach2019blink} uses such switches to detect packet retransmissions. However, Blink can only detect large outages on a single path, not congestion and latency differences across multiple paths. Dapper \cite{dapper}, Lossradar~\cite{lossradar}, and In-band Network Telemetry~\cite{INT} provide performance metrics, such as lost packets and queuing delays, but require bidirectional traffic or/and external mechanisms to aggregate performance markings. While Marple~\cite{narayana2017language} could potentially be used to implement performance monitoring, it does not run in today's programmable switches and does not provide flexible rerouting. 

\subsection{Design constraints}
\label{ssec:constraints}

The following constraints drive \name's design:
\begin{enumerate}[leftmargin=18pt]

                \labeledItem[\textbf{R1}] \label{constraintPolicy}\textbf {Respect routing policies:} By default, \name must select amongst equally-preferred routes, replacing arbitrary tie-breaks in BGP, and hot-potato routing. 
				

        \labeledItem[\textbf{R2}] \label{constraintStable} \textbf{Ensure correctness and stability:} \name must prevent loops and oscillatory behavior. 

                \labeledItem[\textbf{R3}] \label{constraintDeployability} \textbf{Deployability:} \name should not require any coordination between ASes. A single AS deploying \name should also benefit from it without upgrading its entire network.

        \labeledItem[\textbf{R4}] \label{constraintOneDirection} \textbf{Support asymmetric routing:} Due to asymmetric routing, a \name switch may not see both directions of traffic, it must, therefore, be able to estimate and improve performance from one-way traffic. 
    

        \labeledItem[\textbf{R5}] \label{constraintFlowAffinity} \textbf {Respect flow affinity:} To avoid performance degradation due to reordering of packets that could result from sending packets of the same flow across different paths, \name must enforce flow-path affinity.

                \labeledItem[\textbf{R6}] \label{constraintHardware} \textbf {Fit today's switches:} \name should fit within the scarce memory (dozens of MB at best~\cite{Jin:2017:NBK:3132747.3132764}), restricted operations set (e.g., no floating points) and parallel memory accesses available to existing programmable network hardware.

                \labeledItem[\textbf{R7}] \label{constraintBw} \textbf {Limit bandwidth usage:} \name We must limit bandwidth usage between the data and control planes, regardless of the traffic rate and burstness.

\end{enumerate}

\remove{
\subsection{Constraints}
\begin{itemize}
\item One direction of traffic \& one point (a)
\item few programmable switches (possibly just one)
\item per-flow routing
\item single point of observation  (everlow, lossradar)
\item no control over end-hosts to monitor/connection management 
(faceboook, google)
\item limited Data plane flexibility 
\item limited monitoring bandwidth to allow monitoring multiple prefixes (evrflow/ corescope)
\item policy-compliant (equivalent cost) paths only
\end{itemize}

}

%% file: overview.tex
\section{Overview}
\label{sec:overview}

\begin{figure}
\centering
\includegraphics[width=0.8\columnwidth]{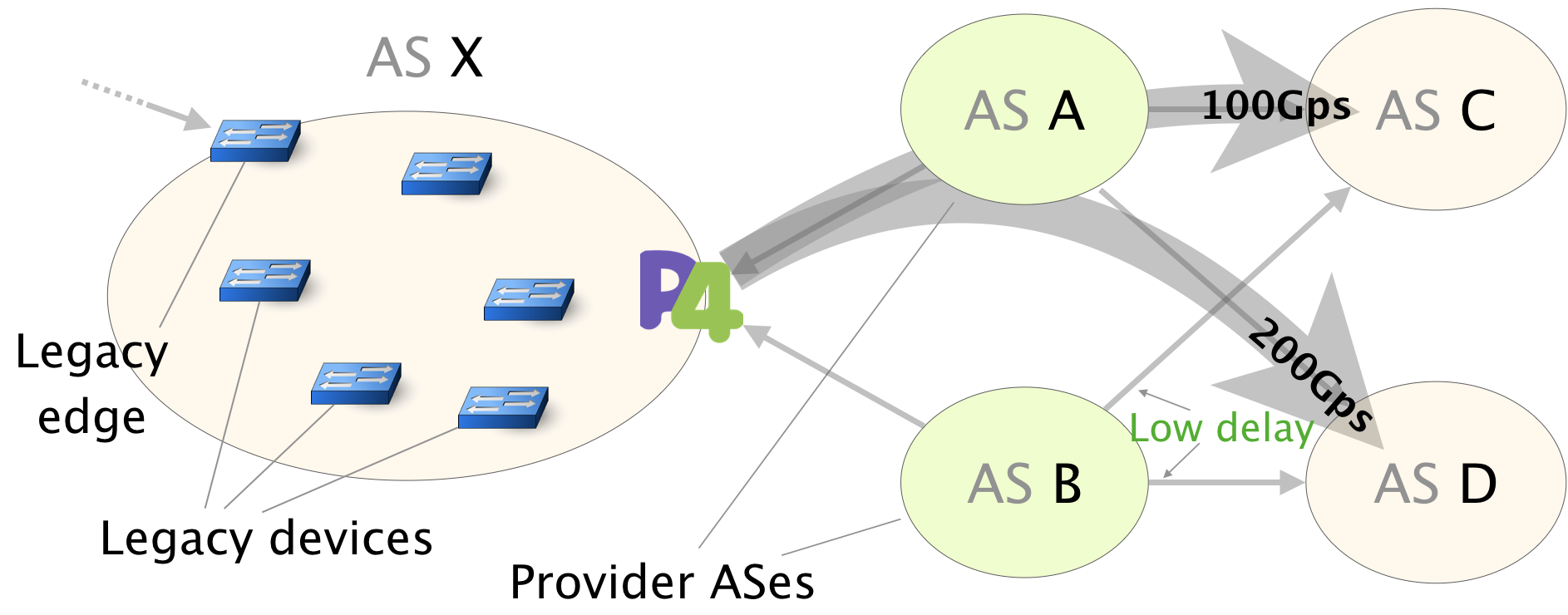}
\caption{$ASA$ and $ASB$ are providers for the other three ASes. $ASX$ has several legacy switches and a \name-capable switch; not all edge switches in $ASX$ run \name.}
\label{fig:overrview2}
\end{figure}


\name is a closed-loop control system that dynamically adapts how a stub AS
forwards its outgoing traffic across multiple policy-compliant routes according
to observed performance and operators objectives.

We illustrate \name operations on a simple running example (Fig.~\ref{fig:overrview2}) in which a stub network, $ASX$, routes traffic to multiple destinations among which $ASC$ and $ASD$. 
$ASX$ knows two equally-preferred paths to reach both destinations through its providers, $ASA$ and $ASB$, with whom $ASX$ has $250$~Gbps links. BGP's arbitrary tie-breaking selects $ASA$ as the next-hop for traffic to $ASC$ and $ASB$ for traffic to $ASD$.
Unbeknownst to $ASX$, the path via $ASB$ has a much lower delay to $ASC$ and a slightly lower delay to $ASD$. Only one (edge) devices of $ASX$ is programmable (\ref{constraintDeployability}).

\myitem{Inputs} To use \name, the operator first specifies the \textbf{prefixes} of interest\footnote{few hundreds (in expectation) accounting for most of the traffic~\cite{sarrar2012leveraging,fang1999inter}}, together with their typical traffic \textbf{demands}.\footnote{adequately accurate estimates, are easy to obtain~\S\ref{ssec:inputs}.} In our example, $ASX$'s operator wants \name to optimize for destinations $ASC$ and $ASD$, which drive $100$ and $200$ Gbps of traffic respectively.
Then, the operator specifies her \textbf{objectives} which in our example are (a) to minimize the delay to both destinations; and (b) to load balance traffic across the next-hops, as long as delay is not increased by >10\%. 
Note that \name automatically learns the policy-compliant next-hops from BGP (\ref{constraintPolicy}).


\vspace{.05in}
\myitem{System} To satisfy the operator's objectives, \name implements a control loop which\ldots
\vspace{-.05in}
\begin{itemize}[leftmargin=12pt]
        \setlength\itemsep{-3pt}
        \item[] \ldots~directs traffic to alternative next-hops
        \item[] \ldots~monitors performance across prefix-nexthop pairs
        \item[] \ldots~computes an optimized traffic allocation to next-hops
        \item[] \ldots~actuates appropriate traffic shifts in the data plane
\end{itemize}

\begin{figure}[t]
\centering
\includegraphics[width=\columnwidth]{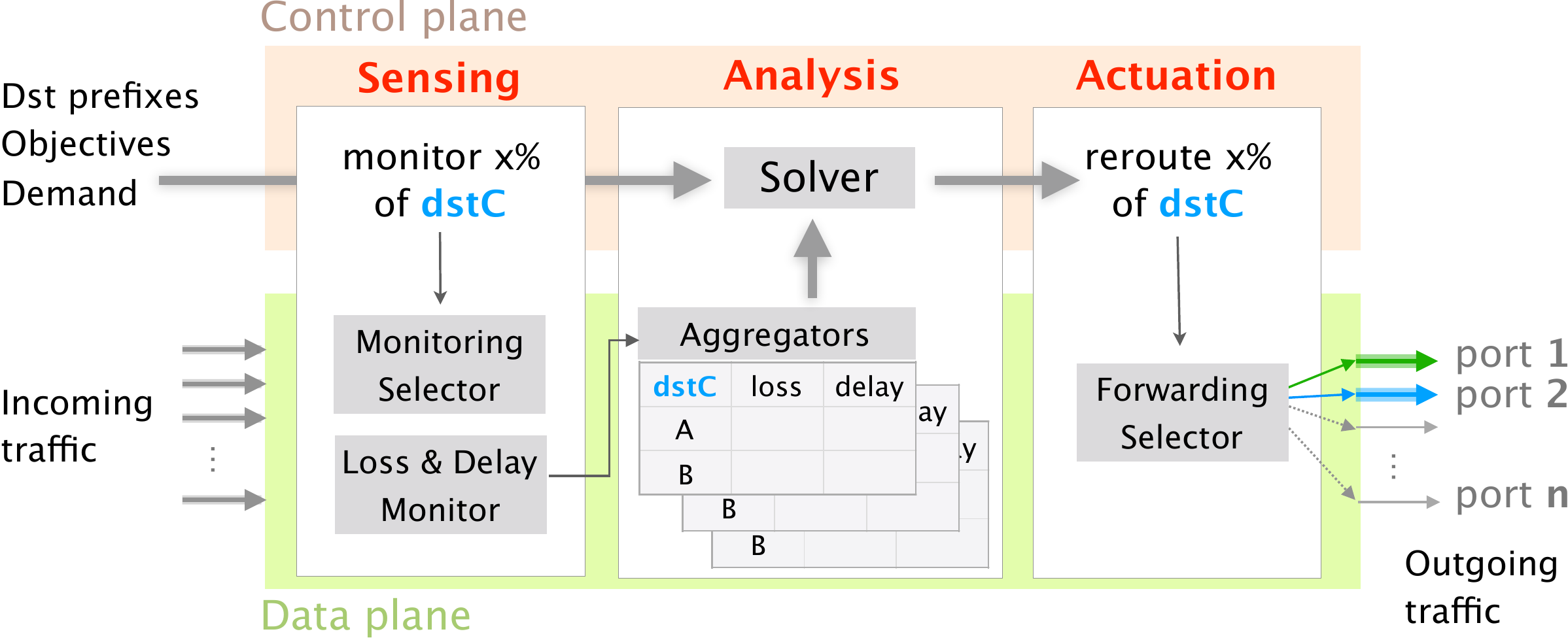}\vspace{-.05in}
\caption{\name is a closed-loop control system with sensing, analysis, actuation split across data and control planes.}
\label{fig:overrview_new}
\end{figure}

\noindent \name splits the above functions across its control- and data-planes (Fig~\ref{fig:overrview_new}).
The data-plane collects and aggregates measurements for the control-plane to analyze (\textbf{sensing}). The control-plane decides which traffic to monitor and which traffic to reroute to which next-hops (\textbf{analysis}). The data-plane receives and enforces these decisions (\textbf{actuation}). 
\name sensing and actuation operates at the granularity of a ``slot'', which we define as a small amount of traffic to a particular prefix. Operating at a per-slot granularity provides measurement efficiency, improved stability and better resource utilization. For instance, slot-based routing enables \name to use paths that can not support all the traffic for a given prefix.

Coming back to our example, $ASD$ receives twice the traffic as $ASC$. Assuming
a total of $3$,$000$ slots, \name allocates $1$,$000$ slots to $ASC$, and
$2$,$000$ slots to $ASD$, with each slot carrying around $0.1$~Gbps of traffic.

\myitem{Data plane:} \name data plane enforces the per-slot monitoring and forwarding decisions made by the control plane. To scalably monitor effectively satisfying \ref{constraintHardware}, \name exploits TCP's semantics together with probabilistic data structures to analyze the relevant packets, aggregate the measurements(\ref{constraintBw}), and actuate the corresponding forwarding decisions (\S\ref{sec:dataplane}). Note that, while \name relies on TCP, it only requires some TCP flows to exist per prefix, meaning it can still be useful even in QUIC-dominated Internet.
To flexibly forward, \name uses two match-action tables and a novel memory mapping scheme (\S\ref{ssec:fwd_monitoring_tables}), that allows it to seamlessly adapt to BGP updates, prefix or policy changes, consistently satisfying \ref{constraintPolicy}





In our example, \name reroutes $1$ slot of traffic to each destination via the alternative next-hop, namely $ASB$ (as decided by the control plane) and monitors $4$ slots one for each destination, next-hop pair. As a result, aggregated loss and delay measurement for each pair will be available to the control plane.

\myitem{Control plane:} \name control plane pulls aggregated data plane measurements, and computes a new forwarding state based on these and the operator objectives (\S\ref{ssec:solver}) by formulating and solving a linear optimization program(\S\ref{ssec:solver}). 

The main challenge in computing a new forwarding state is the conflicting objectives that the operators often have. In our example, the operator wants low delay (primary) and balanced load (secondary). These cannot be satisfied together as $ASB$ offers lower delay for both destinations. This is a deliberately simple example: since performance for $ASC$ improves more, $ASD$'s traffic should be load balanced. But the problem becomes more complex as the number of prefixes, next-hops, and objectives grows. 


%
 
\name moves to the computed forwarding state on a slot-by-slot basis while tracking and reactive any performance degradation to avoid heavily congesting remote bottlenecks potentially violating \ref{constraintStable}. Slot-by-slot traffic shifts also reduce the risk of oscillations, even when multiple \name systems co-exist, by adding randomness and therefore avoiding synchronization~\cite{gao2006avoiding}

%% file: data.tex
\section{\name Data Plane}
\label{sec:dataplane}

\name's data plane uses compact data structures and efficient algorithms to flexibly forward traffic (\S\ref{ssec:fwd_monitoring_tables}) and accurately measure delay (\S\ref{subsec:delay}) and loss (\S\ref{ssec:loss_component}). We also discuss the impact of adversarial inputs and defenses (\S\ref{ssec:adversarial}).

\subsection{Selector stage}
\label{ssec:fwd_monitoring_tables}



The \selector enforces the forwarding and monitoring decisions communicated by
the control plane (\S\ref{sec:overview}) on a per-prefix basis. The forwarding
decisions correspond to the number of slots to forward to given next hops,
while the monitoring decisions correspond to the number of slots to collect
statistics for on given next hops.

\begin{figure}[t]
\centering
\includegraphics[width=\columnwidth]{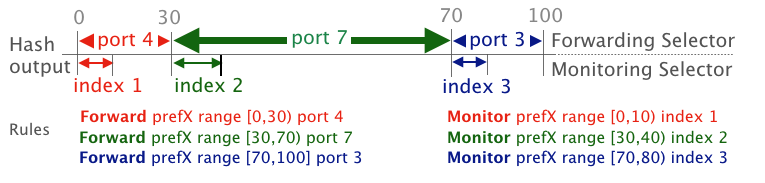}
\vspace{-.1in}
\caption{\name uses two match-action tables to flexibly forward the traffic according to the control-plane decisions and flexibly monitor a given fraction of traffic per next hop. 
}
\label{fig:tables}
\end{figure}

The \selector implements slot-based
forwarding and monitoring by first hashing each incoming packet to a range
[$0$, $k$] and then using two match-action tables to identify sub-ranges [$i$, $j$) of
of the range [$0$, $k$] that need to be to be monitored or forwarded to a given port. The two tables, \fwd and \mnt, use the same type of keys composed of: (i) a prefix; and (ii) a range [$i$, $j$) which identifies a subset of traffic. In the \fwd table, each key maps to a next hop. In the \mnt table, each key maps to the index of a memory block of a table (\agg (\S\ref{subsec:delay}-\ref{ssec:loss_component})) in which the corresponding aggregated statistics will be stored. By adapting the contents of each table, the controller can flexibly adapt the forwarding and monitoring behavior.

\myitem{Example:} Fig.~\ref{fig:tables} shows an example with a
hash range of $0$-$100$, and 3 rules in each table. The rules are such that, in expectation, 30\% of packets (subrange $0$--$30$) to prefix `prefX' will be forwarded to port $4$. Additionally, 1/3 of these packets (subrange $0$--$10$) will be monitored before being forwarded, with the monitoring results stored in index $1$ of the \agg. 
Observe that the flexible design of the \mnt table allow seamless adaptation to the system's dynamics. For example if the BGP peer at port 4 withdraws prefX, then the range of the green (second) rule in the \fwd could be expanded to
include hash outputs $0$-$30$, and the red (first) rules in both the \fwd and \mnt will
be deleted. The index 1 of the \agg used to store measurements for this prefix-nexthop pair can also be reset and assigned to another one.

\subsection{Measuring delays}
\label{subsec:delay}

This component is responsible for accurately and scalably measuring the delay of any flow belonging to one of the monitoring slots enforced by the \selector.
It relies on a \mon and an \agg. The \mon estimates the delay observed by each flow by tracking specific TCP metadata, while the \agg accumulates these statistics which are eventually pulled by the control plane. 

\myitem{Estimating delay:} 
To estimate the delay of a given flow in the presence of asymmetric routing, the \delm computes the time elapsed between its TCP SYN and the first ACK (similarly to \cite{synack}). While doing so means that \name only measures delay at connection setup, it also minimizes the noise from application-level effects which are likely to be more significant for later packets.


\remove{
\begin{figure*}
 \centering
 \begin{subfigure}[t]{.185\textwidth}   \includegraphics[width=\textwidth]{figures/delay1.png}
  \caption{}
\label{fig:mdelay1}
 \end{subfigure}
 \qquad
  \begin{subfigure}[t]{.18\textwidth}   \includegraphics[width=\textwidth]{figures/delay2.png}
  \caption{}
  \label{fig:mdelay2}
 \end{subfigure}
  \qquad
  \begin{subfigure}[t]{.18\textwidth}   \includegraphics[width=\textwidth]{figures/delay3.png}
  \caption{}
  \label{fig:mdelay3}
 \end{subfigure}
  \qquad
  \begin{subfigure}[t]{.28\textwidth}   \includegraphics[width=\textwidth]{figures/delay4.png}
  \caption{}
  \label{fig:mdelay4}
 \end{subfigure}
\vspace{-.1in}

\label{fig:delayEstimation}
\vspace{-.1in}
\end{figure*}
}

\begin{figure}[t]
 \centering
 \begin{subfigure}[b]{.275\columnwidth}
 \includegraphics[width=\textwidth]{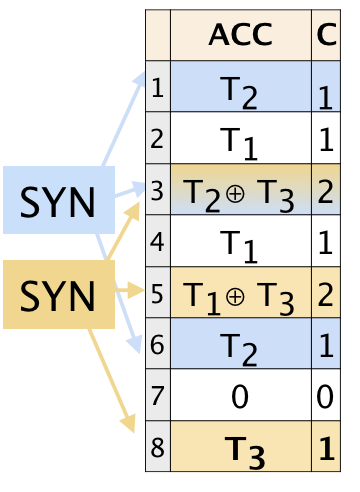}
  \caption{}
\label{fig:mdelay2}
 \end{subfigure}
 \begin{subfigure}[b]{.275\columnwidth}
 \includegraphics[width=\textwidth]{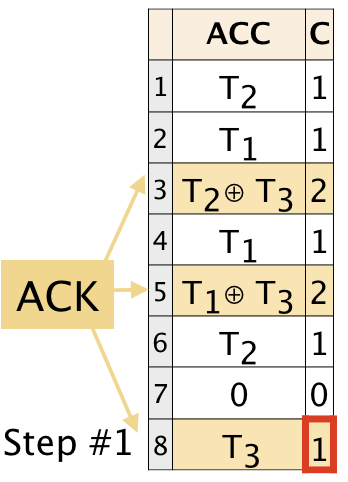}
  \caption{}
  \label{fig:mdelay3}
 \end{subfigure}
 \begin{subfigure}[b]{.43\columnwidth}
 \includegraphics[width=\textwidth]{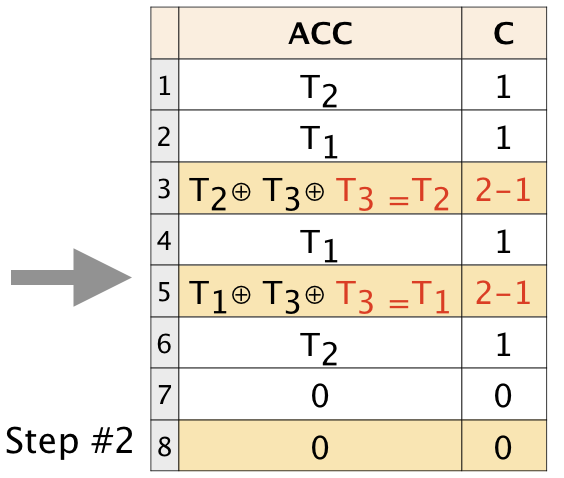}
  \caption{}
  \label{fig:mdelay4}
 \end{subfigure}
 \caption{\delm:(a) SYNs of different flows (blue/above \& yellow/below) increment different indexes; (b) The first ACK of the yellow flow checks that all its indexes (3,5,8) are set, and reads the  timestamp of the yellow SYN from the reversible index 8; (c) The same ACK removes the footprint of the yellow flow by XOR-ing T3 to the indexes of (3,5,8), and decrementing their counters.}
 \label{fig:delayEstimationN}
\end{figure}

Recording timestamps at scale is challenging. Indeed, simply storing the SYN timestamp and the 5-tuple in a hash table does not scale since it requires >100 bits per measurement. 
To address this problem, we use a combination of two probabilistic data structures: an \acc, for storing sums of timestamps at each index, and a \coun for counting how many timestamps are in each sum in the \acc. In essence, the \coun can be seen as a Counting Bloom Filter~\cite{cbf}, while the \acc is similar to an Invertible Bloom Lookup Table~\cite{IBF}. We use XOR ($\xor$) as sum operator rather than a simple addition --- while both $+$ and $\xor$ are recoverable (given $A$ and $A \xor B$ or $A + B$, one can recover $B$), $\xor$ cannot cause overflows. Unlike previous works~\cite{flowradar,lossradar} that send their full Bloom filters to the controller to be decoded (incurring both compute and bandwidth expense), we measure entirely in the data plane, and only expose aggregated statistics to the control plane which can pull them asynchronously. 


\myitem{Example, Fig.~\ref{fig:delayEstimationN}:} As SYNs of different flows arrive (Fig.~\ref{fig:mdelay2}), we hash their 5-tuples with multiple hash functions, thus generating multiple indexes. Here the yellow (lower) flow is hashed to $(3,5,8)$, and the blue (upper) flow to $(1,3,6)$. Each entry of the \acc in those indexes is $\xor$-ed with the timestamp of the SYN. Additionally, the \coun of each entry is incremented. Different SYNs can end updating the same index, \eg index $3$ in Fig.~\ref{fig:mdelay2}. 


On receiving an ACK, we first compute the corresponding indexes using the same hash functions. If all the corresponding \coun values are non-zero, then we know that the SYN timestamp is contained in the \coun. In Fig.~\ref{fig:mdelay3}, the ACK of the yellow flow arrives and finds its indexes set. To get the timestamp of its corresponding SYN, we need to find one index among the indexes to which the ACK is hashed, whose value in the \coun is one. We will call this index reversible.
The same index in the \acc yields the timestamp for this flow's SYN, thus allowing us to compute its delay. In Fig.~\ref{fig:mdelay3}, the ACK finds a value equal to $1$ in the index $8$, namely the third of the three indexes it is hashed to. Thus, the timestamp of the SYN is at index $8$ in the \acc.

To erase the footprint of a SYN from the \delm, we decrement each of the hashed indexes in the \coun, and $\xor$ the
recovered timestamp with the sums at these indexes in the \acc. In Fig.~\ref{fig:mdelay4}, we illustrate the result of this process; observe that by $\xor$-ing the timestamp in each of the hashed indexes, the effect of the yellow SYN vanishes.

\myitem{Keeping the \delm healthy:} 
In the common case, the \delm stores some per-flow state only during the handshake as an ACK removes the memory footprint created by the corresponding SYN. This allows the \delm to scale with the number of flows regardless of their rate and duration. Still, a large number of SYNs not followed by corresponding ACKs can pollute the \delm. This challenge can be easily addressed by keeping track of the number of SYNs in the \delm and not add new ones if the filter has exceeded its capacity (number of elements it can store based on allocated memory, \S\ref{ssec:delay_eval}). Alternatively, the filter can be reset periodically.

\myitem{Aggregating statistics:} The \agg stores the delay measurements per prefix-nexthop pair in an array with two values per index: one for storing the sum of the delays and one for storing the number of delay measurements contained in the former. The control plane can pull the measurements for a prefix-nexthop pair or for all pairs at once, and calculate the mean delay. For example, in Fig.~\ref{fig:mdelay4}, once the ACK has read the timestamp of its SYN it calculates the time elapsed since then and update the values in the index that is mapped to its prefix and output port. 
The mapping between the prefix-nexthop pair and the index in the \agg is assigned by the control plane and communicated via the \mnt. Thus, to monitor different prefixes or different number of next hops for some prefixes, one just changes this mapping instead of re-allocating memory and needing recompilation (see example in \S\ref{ssec:fwd_monitoring_tables}).



\subsection{Measuring loss rates}
\label{ssec:loss_component}

The design and challenges of the loss measurement component are similar to those for delay with some key distinctions. In particular,
to measure the loss rate, the \mon tracks the amount of re-transmitted and regular packets, while the \agg accumulates the counts for each category.

\begin{figure}[t]
\centering
\includegraphics[width=\columnwidth]{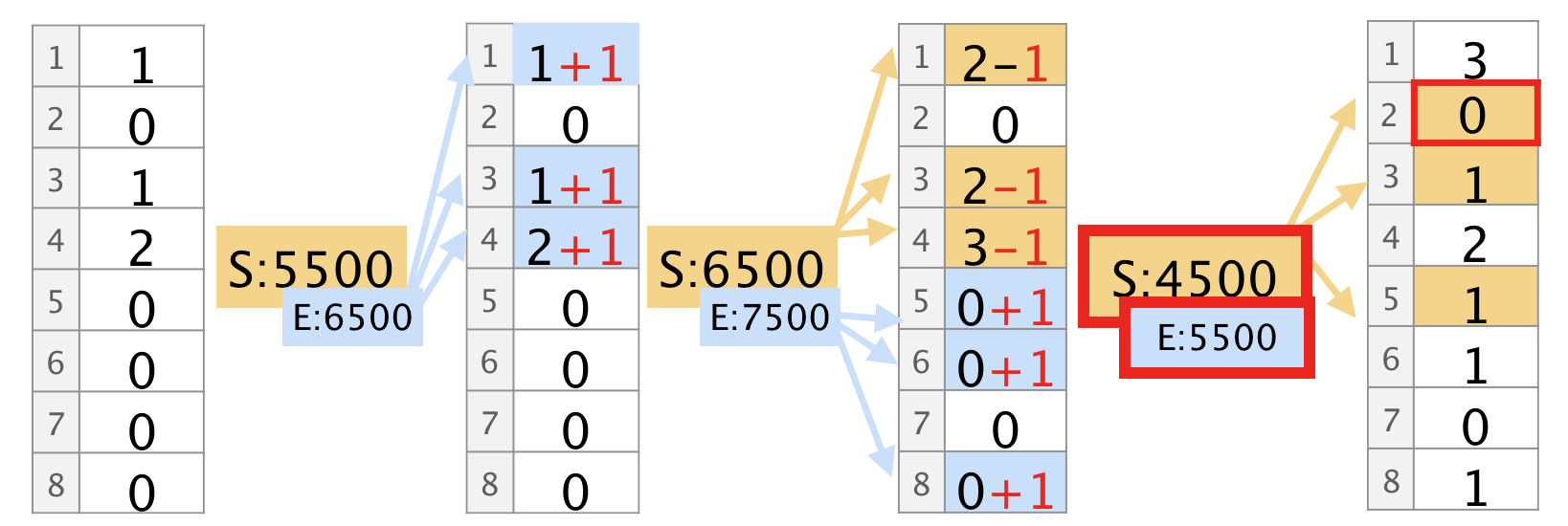}
\vspace{-.2in}
        \caption{Here the \losm sees three packet arrivals, $2$ in-order and $1$ retransmit. The first, with sequence number S:5500 has the next expected sequence number E:6500, and inserts the latter into the CBF by incrementing the indexes corresponding to the E:6500 (blue indexes, 1, 3, and 4). The second packets finds its indexes (now yellow, 1, 3, and 4) non-zero, thus knows it was expected. It cleans itself out, and inserts the next expected packet (blue indexes, lower) . The third one, a retransmit, finds one of its indexes ($2$) unset.}
\label{fig:los_example}
\end{figure}

\myitem{Estimating loss rate:} Measuring retransmissions at scale is challenging since one cannot simply store every packet and compare new arrivals against the past history to identify duplicates. Our solution, somewhat surprisingly, requires only a few bits per flow, at the cost of one minor compromise: the inability to distinguish reordering from retransmissions. Given that reordering also hurts TCP~\cite{blanton2002making},  mistakenly accounting for it as loss is not a significant downside, if it is one at all.

Our solution keeps only one element per flow by exploiting TCP semantics and the fact that, given a TCP packet $p$, one can compute the \emph{next} expected sequence number based on $p$ sequence number and payload length. By storing this expected sequence number, we can check whether the next packet is either a retransmission or an out-of-order packet. 
Instead of storing a $16$-bit (expected) sequence number, $e$, we can insert it into a counting bloom filter (CBF), \ie the same data structure as our \coun for delay estimation. Since, packets across flows can share sequence numbers, we insert the concatenation of the $5$-tuple with the sequence number instead. Increasing the length of the inserted value is immaterial, as the length of the hash output is the same.

Whenever a packet with sequence number $s$ arrives, we check the CBF for <$5$-tuple, $s$>:  If the entry does not exist, the packet is out-of-order or a retransmit. If the entry exists, the packet is in-order and we delete it from our filter by decrementing all the indexes <$5$-tuple, $s$> hashes to. We then insert the \emph{next} expected packet by incrementing all the indexes that <$5$-tuple, $s+tcp.len$> hashes to.

Not all packets carry information regarding previous segments. For instance, an ACK that does not carry any TCP data will be followed by a packet of the same sequence number regardless of whether the former was lost or not. Similarly, KEEPALIVE messages (commonly used in Web traffic) contain an ``unexpected'' sequence number: one byte less than the previously sent sequence number. To avoid these issues, we only use packets with TCP payload. This does not disrupt functionality, as for every non-zero-payload packet whose subsequent sequence number we store, there will be a non-zero-payload packet that can remove it, even if it comes after multiple zero-payload ACKS.

\myitem{Example, Fig.~\ref{fig:los_example}:} In this example, we illustrate how $3$ packets (the last one being a retransmitted one) of a flow update the CBF. The yellow (upper) box contains their sequence number, and the blue box (lower), the sequence number of the expected packet. The first packet inserts the fingerprint of the expected (second) one by incrementing the values stored in the indexes that the expected sequence number (concatenated with the 5-tuple of the flow) hashes to (blue indexes). Thus, when the second packet arrives, it will find all hashed indexes of its sequence number set (yellow indexes), and will consider itself expected. This is not true for the third packet whose indexes are not all set and is a retransmit.

\myitem{Keeping the \mon healthy:} 
Similarly to the \delm, the \losm contains one item per flow regardless of its rate as the structure ``cleans itself'' with incoming packets. 
In particular, once a flow terminates, the corresponding RST or a FIN removes the flow permanently. Still, out-of-order and lost packets will, in most cases, cause some packets to stay in the filter. However, this represents a very small fraction of packets, as we discuss in \S\ref{ssec:loss_eval}. To avoid overflowing the \mon, a counter in the data plane can keep track of the number of flows using it. If the filter's capacity is exceeded, insertions are stalled until some of the flows terminate. Alternatively, the filter can be reset periodically as we show in \S\ref{ssec:loss_eval}.

\myitem{Aggregating statistics:} Similarly to the \dela, the \agg stores the number of expected and unexpected packets observed per prefix and next hop.



\subsection{Dealing with adversarial inputs}
\label{ssec:adversarial}

Like any data-driven system, \name is prone to attacks in which malicious end-points or networks aim at faking signals in order to influence its decisions. While possible, and deserving a complete analysis in a follow-up work, we briefly argue why such attacks on \name are hard to perform. 


In order to influence \name's decisions, a malicious end-point could try to: (i) send repeated packets to fake retransmissions; or (ii) send fake pairs of SYNs and ACKs with a small/large timing differences to fool the delay monitor. We note two things. First, such adversarial end points must be hosted within the stub AS, since \name optimizes exit traffic. Assuming basic anti-spoofing techniques are in place (e.g.~\cite{spoof}), each end point has a single IP address to source traffic from. As such, limiting the number of flows tracked per IP would be sufficient to mitigate the attack. Second, \name randomly associates a flow to a next hop, depending on a hash function. As such, the attacker is equally likely to add noise to measurements of \emph{all} next hops, making targeting one next-hop difficult. \name can also defeat attempts to use traceroutes for probing such decisions by randomly forwarding traceroutes to next hops.

Similarly, a malicious transit network can: (i) drop packets to increase the loss rate; or (ii) drop/delay SYNs, SYN/ACK, or ACKs to fool the delay monitor. While this is possible, we note that, by doing so, malicious networks can only make their performance worse, not better. As such, malicious networks can only push away traffic, not attract more.
Observe, that an attacker cannot craft a SYN/ACK packet for every SYN it receives to fake low latency as she does not know the sequence number that the receiver will use until the actual SYN/ACK packet is received. 

Finally, attackers can also attempt to pollute \name's data structures. An efficient way to mitigate such pollution is to periodically reset the data structures, as we discuss in \S\ref{sec:evaluation}.


%% file: implementation.tex
\section{Hardware Design}
\label{sec:tofino_implementation}

Our design needs modification to fit a real Protocol Independent Switch
Architecture (PISA) switch. We briefly explain the key constraints imposed by
PISA and how we adapted the Delay and Loss monitors accordingly. We have tested
our design in a Barefoot Tofino Wedge 100BF-32X.

\myitem{PISA constraints:}
A packet traversing a PISA switch goes through a pipeline of stages.
Besides the limited memory and instruction set, which our design already addresses, there are constraints on the sequence of memory accesses~\cite{Ben_Basat_2018,snappy18}.
First, a packet cannot read or write multiple memory addresses in the same memory block. Second, memory blocks are tied to a single stage in the pipeline and can only be accessed in it. This is to avoid contention
from stages processing different packets simultaneously. 
Similarly, accessing stages in a different order or multiple times per packet is not possible.

\begin{figure}[t]
 \centering
 \begin{subfigure}[b]{.34\columnwidth} 	\includegraphics[width=\textwidth]{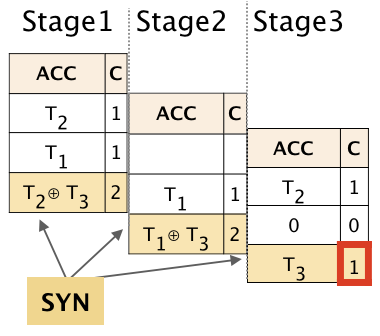}
	\caption{}
	\label{fig:tofDelay}
 \end{subfigure}
 \begin{subfigure}[b]{.64\columnwidth} 	\includegraphics[width=\textwidth]{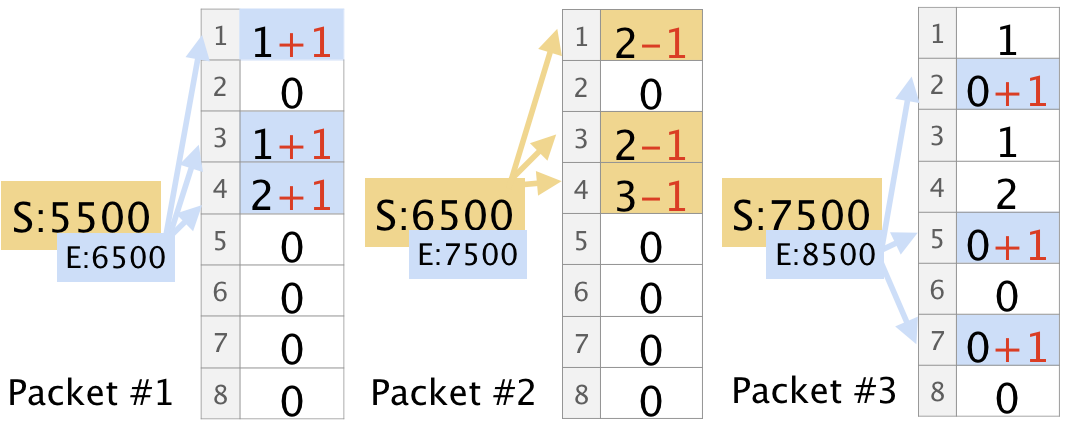}
	\caption{}
	\label{fig:tofLoss}
 \end{subfigure}
 \caption{(a) We implement the \delm as a series of arrays; (b) A packet can either check if it is expected or insert the next expected packet in the \losm.}
 \label{fig:tof}
\end{figure}

\myitem{Delay Monitor modifications:}
To access any kind of Bloom Filter, including those in the Delay Monitor, we need to access multiple indexes, each corresponding to the output of a hash. For instance, in Fig.~\ref{fig:mdelay2}, the yellow SYN would need to access three indexes corresponding to the yellow indexes. In PISA though, one cannot
concurrently access multiple indexes of the same memory block.
We thus divide the two tables of the \mon into smaller chunks, and constrain each hash to index a single chunk as seen in Fig.~\ref{fig:tofDelay}. Now, chunks reside in different stages of the pipeline and can be accessed serially.

Serializing accesses creates another issue. Particularly, when an ACK arrives, the \mon first needs to find out if it corresponds to the first ACK of a flow whose SYN is in the \acc (Fig.~\ref{fig:mdelay3}), and if so, decrement all corresponding indexes in the \coun.
For this, the SYN will need to traverse all three pipeline stages in Fig.~\ref{fig:tofDelay} to check whether all corresponding indexes of the \coun are non-zero.
But after doing so, the packet cannot
return to stage $1$ and decrease their values in the \coun. To address this, the monitor recirculates packets
corresponding to first ACKs.
Observe that even if we could rely on SYNACK, which is impractical due to asymmetric routing, we would still not be able to avoid recirculation. Indeed, even if an incoming ACK knew upon arrival that the timestamp of the corresponding SYN is in the structure, it will still need to find a reversible index to read this timestamp and then $\xor$ it to all (previous) stages.
As an illustration, in Fig.~\ref{fig:tofDelay}, the reversible index is in stage $3$. At the time the packet reads it, it can no longer return to stages $1$ and $2$, and  $\xor$ it to the corresponding indexes.

\myitem{Loss Monitor modifications:} Similarly here, we need to split the CBF into multiple chunks and stages. Recall that every incoming packet needs to check if it is expected, remove itself, and insert the next expected packet in the CBF. This results in two violations of the PISA constraints.

First, a packet needs to access each memory chunk (in each stage) in two different indexes, one corresponding to the output of itself, whose  value  it needs to decrement, and one corresponding to the next expected packet, whose  value it needs to increase.
Second, the former access is conditioned on whether the packet is expected or a re-transmission something which will only be known after the packet traversed all stages.

To address the first violation, we allow each packet one of the two operations, either to remove itself, if it is expected, or to insert the next expected one iteratively. To achieve this, we keep track of the number of packets seen by each flow. Particularly, when a packet arrives, it checks the number of  non-zero-payload packets its flow has already sent. If this number is even, as for S:5500 and S:7500 in Fig.~\ref{fig:tofLoss}, then the packet will insert the next expected one in the CBF. If the number is odd, as for S:6500 in Fig.~\ref{fig:tofLoss}, the packet will try to find its footprint in the CBF and remove it. We use a counting bloom filter to efficiently keep track of the number of packets.

To address the second violation, we assume all packets to be expected and recirculate packets that violate this assumption. In more detail, on arrival, a packet whose flow has sent an odd number of packets reads and decrements the indexes corresponding to it in the CBF. If the packet was indeed expected, \ie all read values are non-zero (as for S:6500 in Fig.~\ref{fig:tofLoss}), the packet increments the \acc and leaves the device. If the packet was a retransmission, it is recirculated to re-increment the indexes it wrongly decremented.

%% file: multi_control.tex
\section{\name Control Plane}
\label{sec:alt}

In this section, we describe \name's control plane and how it leverages
measurements from the data plane to improve forwarding decisions. We start by
describing the control-plane inputs (\S\ref{ssec:inputs}). We then explain how
it solves the induced optimization problem (\S\ref{ssec:solver}).

We describe the simplest version of the control plane that would enable performance-driven routing and support conflicting operator objectives. 
To cover additional operational needs, this control plane can be extended for instance to strengthen stability guarantees as shown in \cite{gao2006avoiding}.


  
\subsection{Inputs}
\label{ssec:inputs}



\name triggers the \solver periodically giving as input a description of the environment,
a set of objectives, and optionally, some additional constraints for each prefix,
together with fresh performance statistics. 

\myitem{Environment:} 
The network environment includes  topological, traffic, and routing information.
The former two are provided by the operator and the latter by BGP. 
Topological information corresponds to the set of direct next-hops and their link capacities. Traffic information consists of the set of prefixes that \name should optimize for, together with the volumes they drive. Routing information corresponds to the set of next-hops that \name can use to route each prefix (obtained from routing tables and BGP policies).

Expecting traffic information is reasonable as important prefixes are few and stable over time~\cite{sarrar2012leveraging,fang1999inter}. The traffic volumes to these prefixes can also be estimated accurately~\cite{Sang2000APA,
iqbal2019efficient}. Note that inaccurate traffic volumes won't affect \name's
performance if the direct links are not running at full capacity which
is true in most stub ISPs. If that's not the case, \name might indeed not find the optimal solution but will never deteriorate the performance by moving traffic to a worse next hop.


\myitem{Objectives:} The operator can decide whether they want to: (i) optimize for delay and/or loss; 
(ii) minimize the number of traffic shifts necessary to meet the requirements; 
or (iii) load-balance traffic by minimizing the difference between the most- and the least-used next-hop. Linear combinations of these or similar other objectives are easily implementable.

\name also allows multiple objectives to be flexibly implemented. To do so, the operator needs to express how
important each objective is by defining priorities and how valuable are the
differences among alternative forwarding states by defining tolerance levels.
Objectives with lower priority will only be optimized if there are multiple
equally-preferred solutions, namely solutions that differ from the optimal by no
more than the tolerance level. For example, an operator might want to balance the load across the next-hops, as long as the delay difference between the best- and the used
 next-hop is lower than 10\%. The operator can communicate this to \name by giving a
high priority to delay with $10\%$ tolerance, and a lower priority to load-balancing.

\paragraph{Operational constraints:} \name admits constraints of two types: (i) those that limit the number of next-hops traffic can be spread on; and (ii) those that define performance constraints.
Constraining the maximum number of next-hops per destination might be useful, for instance, to ease debugging. Performance constraints are maximum loss/delay values that traffic for a certain destination should experience. Defining such objectives is useful for meeting Service Level Agreements (SLAs), or particular application requirements.

\myitem{Data plane statistics:}
\name periodically pulls measurements of loss and delay aggregated per prefix and next-hop from the respective \emph{aggregators}. 




\subsection{Solver}
\label{ssec:solver}

The solver is responsible for synthesizing a forwarding state. To do so, it formulates each of the operator's inputs
into a constraint or an objective, creating a linear optimization problem. 
\myitem{Problem statement:} 
Let $N$ be a set of next-hops and $P_r$ the set of destination prefixes to optimize for. 
Let $P_a \subseteq  P_r \times N$ be the set of all pairs of destinations and equally-preferred next-hops (learned by BGP).
The goal is to find a mapping ${F_{t}}: P_a \rightarrow  \mathbb{N}$, namely
the number of slots allocated to each pair (prefix, next-hop) at time $t$ such
that it optimizes the operator's objectives, while adhering to the environmental and operational contraints.
We implement the \solver using Gurobi~\cite{gurobi}.

%% file: dataEval.tex
\section{Evaluation}
\label{sec:evaluation}
\begin{figure*}[t]
 \centering
 \begin{subfigure}[t]{.3\textwidth}   \includegraphics[width=\textwidth]{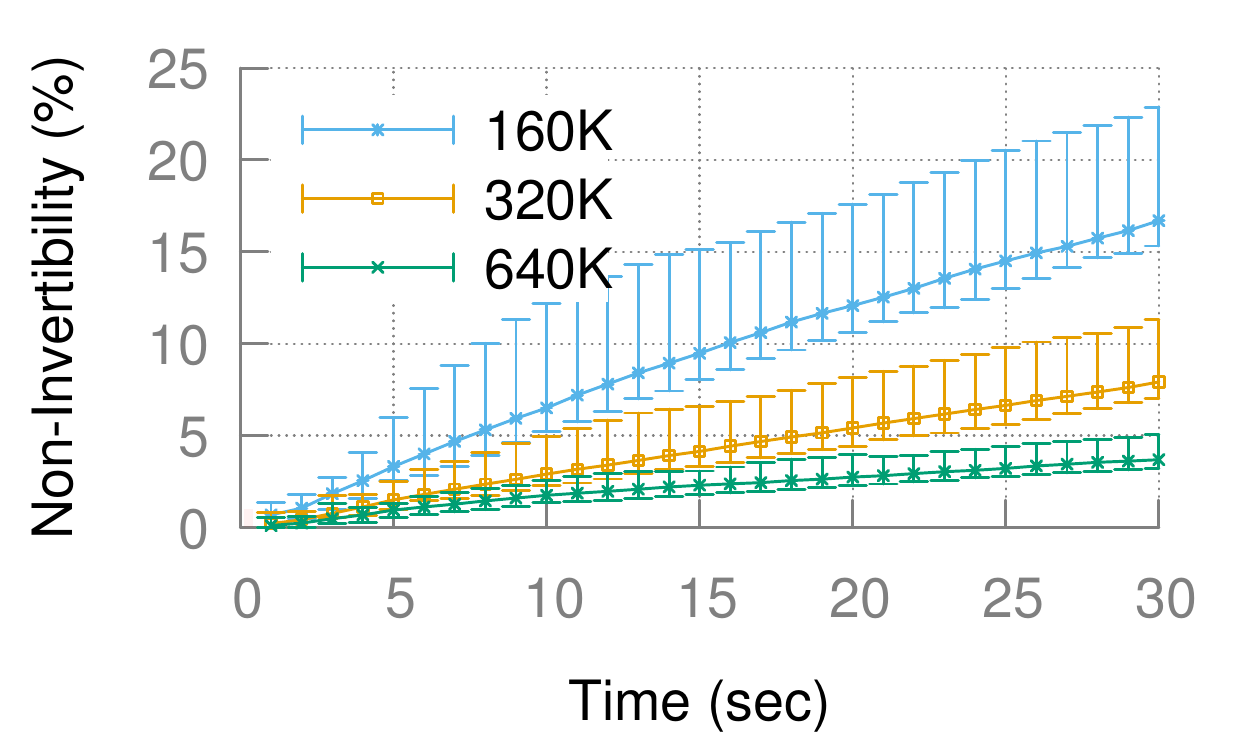}
  \caption{ The probability of an ACK to decode its SYN's timestamp is >95\%  with a 1MB (640K elems) \delm of 2 hashes.
  }
  \label{fig:accdelay}
 \end{subfigure}
 \qquad
  \begin{subfigure}[t]{.3\textwidth}   \includegraphics[width=\textwidth]{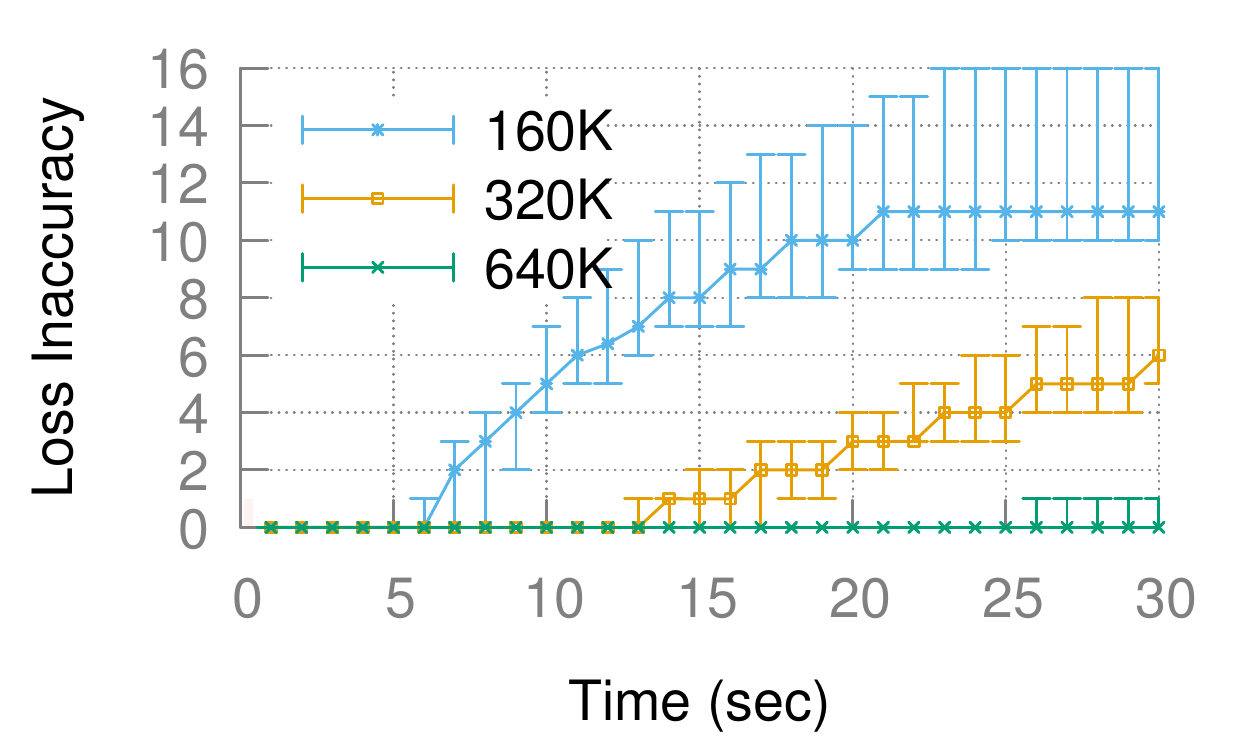}
  \caption{Using 625KB (640K elems) and 2 hashes, the \losm calculates the loss rate with high accuracy.}
  \label{fig:accloss70}
 \end{subfigure}
  \qquad
  \begin{subfigure}[t]{.3\textwidth}
	\begin{tabular}[b]{l l l l l}
	\footnotesize
	\begin{tabular}{@{}l@{}} \emph{Target} \end{tabular}& \begin{tabular}{@{}l@{}} \emph{50th } \end{tabular} &  \begin{tabular}{@{}l@{}} \emph{70th}  \end{tabular} &   \begin{tabular}{@{}l@{}} \emph{95th} \end{tabular}   \\
	\midrule
	\# moves & 0.02 & 0.03  & 0.29 \\
	balance & 0.03 & 0.04 &  0.4\\
	performance & 0.08 & 0.3 &  1.09\\
	combined  & 0.03 & 0.05 &  1.23\\
	\bottomrule
	\vspace{1em}
	\end{tabular}
	\caption{Runtime percentiles in seconds depend on the complexity of the objective.}
      \label{tab:obj}
	 \end{subfigure}
\caption{}
\vspace{-.2in}
\end{figure*}

We evaluate \name's \delm (\S\ref{ssec:delay_eval}), \losm (\S\ref{ssec:loss_eval}) and \solver (\S\ref{ssec:solver_runtime}). For the monitors, we investigate the trade-off between accuracy and memory footprint using real traffic traces and our practical hardware design (\S\ref{sec:tofino_implementation}). 
 We find that, with $1$ MB of memory, the \delm can accurately measure the delay of hundreds of thousands of flows/sec.
Moreover, the \losm can accurately measure loss rate of $36K$ flows/sec with as little as $312$KB of memory.
For the \solver, we focus on runtime, and show that it computes forwarding states for thousands of destinations, across tens of next hops and for various objectives, in less than a second.


\subsection{Methodology}
\label{ssec:datasets}

To evaluate \name's monitors we estimate the memory they use as a function of their accuracy via both theoretical and practical means.
For the theoretical analysis, we assume perfectly behaved TCP traffic (in-order, with expected semantics), with flow rates derived from real traces, and the original design as described in \S\ref{subsec:delay}, \S\ref{ssec:loss_component} with $9$ hash functions\footnote{We chose $9$ following the Bloom Filters heuristic~\cite{BF}.} and without any additional hardware limitations. 
For the practical analysis, we use real traffic traces and our hardware design for Tofino, with only $2$ hash functions\footnote{More engineering effort might allow an implementation of more hashes.}.

For the theoretical analysis, we use two different directions of CAIDA traces (CAIDA.A, CAIDA.B) collected at the Equinix-Chicago monitor in March $2018$~\cite{caida}, and one from MAWI~\cite{mawi} from January $2018$. Together, these contain $\sim$6 billion packets with an average rate ranging from $240$-$3200$~Mbps. For the practical analysis, we use the CAIDA.A trace, which is the noisiest, and feed it to the monitors in $100$ chunks of $30$ seconds. While none of those traces are from a stub network, this has no impact on our analysis, as we are only interested in estimating accuracy and resource usage.

\subsection{\delm}
\label{ssec:delay_eval}

\remove{
\begin{figure}[t]
\centering
\includegraphics[width=.7\columnwidth]{figures/delay.pdf}
\vspace{-.1in}
\caption{\name and tshark measure different delays.}
\vspace{-.1in}
\label{fig:delay_diff}
\end{figure}
}

 \myitem{Accuracy metric:} We calculate the \emph{invertibility}, namely the probability of a successfully computed delay. 
The delay between a SYN and its corresponding ACK can be successfully computed \emph{if} upon arrival of the ACK, there is at least one index that contains \emph{only} the timestamp of the SYN. Other than the memory used, invertibility depends on the number of concurrent delay measurements, the number of hash functions used, and the pollution of the structure due to traffic noise, \eg SYNs that are not followed by ACKs. 
 

\myitem{Theoretical analysis:}
In theory, invertibility is the inverse of the probability of false positive in a regular Bloom Filter: the probability of a SYN being $\xor$ed to indexes which all contain other timestamps is the same as finding all hash outputs set in a regular Bloom Filter during a lookup. 
We calculate the memory requirements for an invertibility of $99.9\%$ (false positive rate in BF of $0.1\%$) using the analytical formula for optimal Bloom Filter design~\cite{BF}. For these calculations, we assume that each handshake completes in <$1$~sec, and that \name needs to monitor all flows in each trace. The results are summarized in Table~\ref{tab:mem}. The \delm would need $12.9$K-$781.5$K elements, corresponding to $6$KB--$381$KB memory assuming an implementation over an array of 16-bit values using 9 hash functions.

\begin{table}[t]
\centering
	\footnotesize
	\def\arraystretch{1.1}
	\setlength{\tabcolsep}{3pt}
	\begin{tabular}{l | l l l | l l l}
	\toprule
	\begin{tabular}{@{}l@{}} \emph{Trace} \end{tabular}&
	\begin{tabular}{@{}l@{}} \emph{SYNs/s } \end{tabular} &
	\begin{tabular}{@{}l@{}} \emph{Elements} \end{tabular}&
	\begin{tabular}{@{}l@{}} \emph{Delay M} \end{tabular}&
    \begin{tabular}{@{}l@{}} \emph{Flows/s} \end{tabular} &
	\begin{tabular}{@{}l@{}} \emph{Elements } \end{tabular}&
	\begin{tabular}{@{}l@{}} \emph{Loss M} \end{tabular}\\
	\midrule
    CAIDA.A & 3.8K   & 54.2K  & 26KB &  36.8K &   529.1K  & 1MB\\
    CAIDA.B &   54.4K &  781.5K &  381KB &  233.8K &  3361.3K  &  6MB\\
    MAWI &  899 &  12.9K  &  6KB &  3.3K &  47.8K &  93KB\\
	\bottomrule
	\end{tabular}
	\vspace{-2pt}
	\caption{ \delm and \losm  would combined need $6.4$M to monitor as many flows/s as there are in the CAIDA.B trace.} 
	\label{tab:mem}
	\vspace{-.05in}
\end{table}

\myitem{Practical analysis:}
        In practice, the filter is gradually polluted by SYNs that are not followed by ACKs. This can happen, \eg under SYN attacks, or when hosts try to reach an offline server. Such noise is common in our traces: in the noisiest trace (which we use for this evaluation), only $40$\% of the SYNs are followed by ACKs. 
        Fig.~\ref{fig:accdelay} shows the median, max, and min non-invertibility probability as a function of time using \{160K, 320K, 640K\} elements in the data structure.
As expected, the failure probability increases with time as the filter gets polluted. Still, \name is very efficient.
Indeed, a \delm with only 320K elements 
 has an invertibility  of >90\%. Another interesting insight is that we can do this with less memory if we periodically reset our \delm, \eg with only 160K elements ($312$KB), we get the same >90\% invertibility if we reset it every $15$ seconds.

\subsection{\losm}
\label{ssec:loss_eval}



        \myitem{Accuracy metric:} We compare the measured loss per flow to its actual loss rate. \name's accuracy is affected by false positives: a retransmitted packet can be considered expected (instead of correctly being assessed as unexpected) and thus not counted towards loss, if all the indexes it hashes to are set. As the \losm is a CBF, its false positive rate depends on the memory and the number of hashes used. 

\myitem{Theoretical analysis:} We use the same method as for the \delm, to calculate memory requirements for achieving a false positive rate of <0.1\%. 
The results are summarized in Table~\ref{tab:mem}. The \losm would need $47.8$K--$3.4$M elements depending on the number of flows/sec in the trace. This corresponds to $93$K--$6$M memory, if the \losm is implemented as an array of 4-bit values with 9 hash functions.

\myitem{Practical analysis:} 
In practice, the \losm's accuracy is deteriorated by three more factors.
First, out-of-order packets are not only classified as losses, but also pollute the structure as explained in \S\ref{ssec:loss_component}.
Second, flows terminating unexpectedly (\ie without FIN/RST) remain in the monitor until it is reset, decreasing its effective capacity. 
Third, the \losm can miss some loss events due to the compromise for PISA constraints: it only checks whether every other non-zero-payload packet has the right sequence number.

        Despite these impairments, \name is in practice, very accurate. Fig.~\ref{fig:accloss70} shows the (max, min, and median across all runs) $70$\textsuperscript{th} percentile of difference across all flows between their estimated loss rate and the ground truth reported by tshark. We plot $70$\textsuperscript{th} as lower percentiles have zero error, and thus unsuitable for studying the memory trade-off. We find that a \losm with only 640K elements (625KB  assuming 4bits/element) is almost perfect for 30 sec. Like the \delm, resetting every 15~sec would allow smaller implementations to be similarly accurate.

%% file: control_eval.tex
\subsection{\Solver runtime}
\label{ssec:solver_runtime}
We investigate the influence of each parameter of the operational
environment (\S\ref{ssec:solver}) on the \Solver's runtime. 


\begin{figure*}[t]
 \centering
 \begin{subfigure}[t]{.6\columnwidth}   \includegraphics[width=\textwidth]{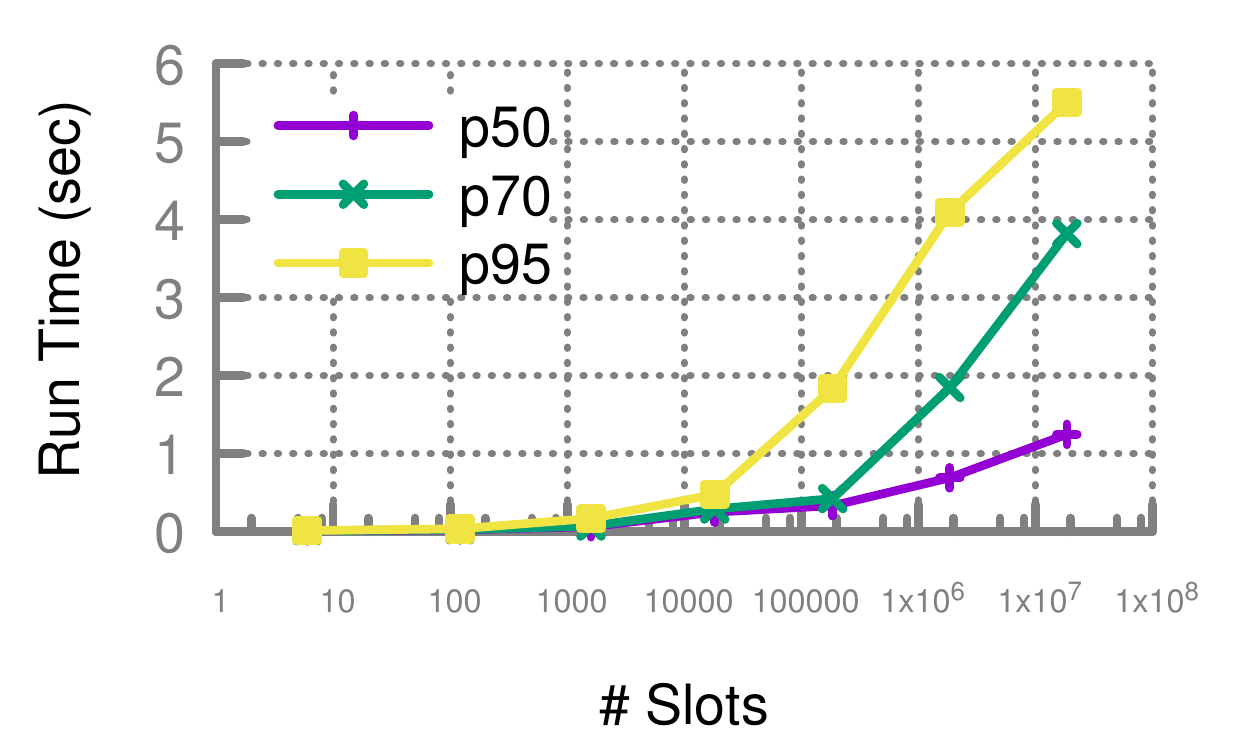}
   \vspace{-.2in}
  \label{fig:lp_demand}
 \end{subfigure}
 \begin{subfigure}[t]{.6\columnwidth}   \includegraphics[width=\textwidth]{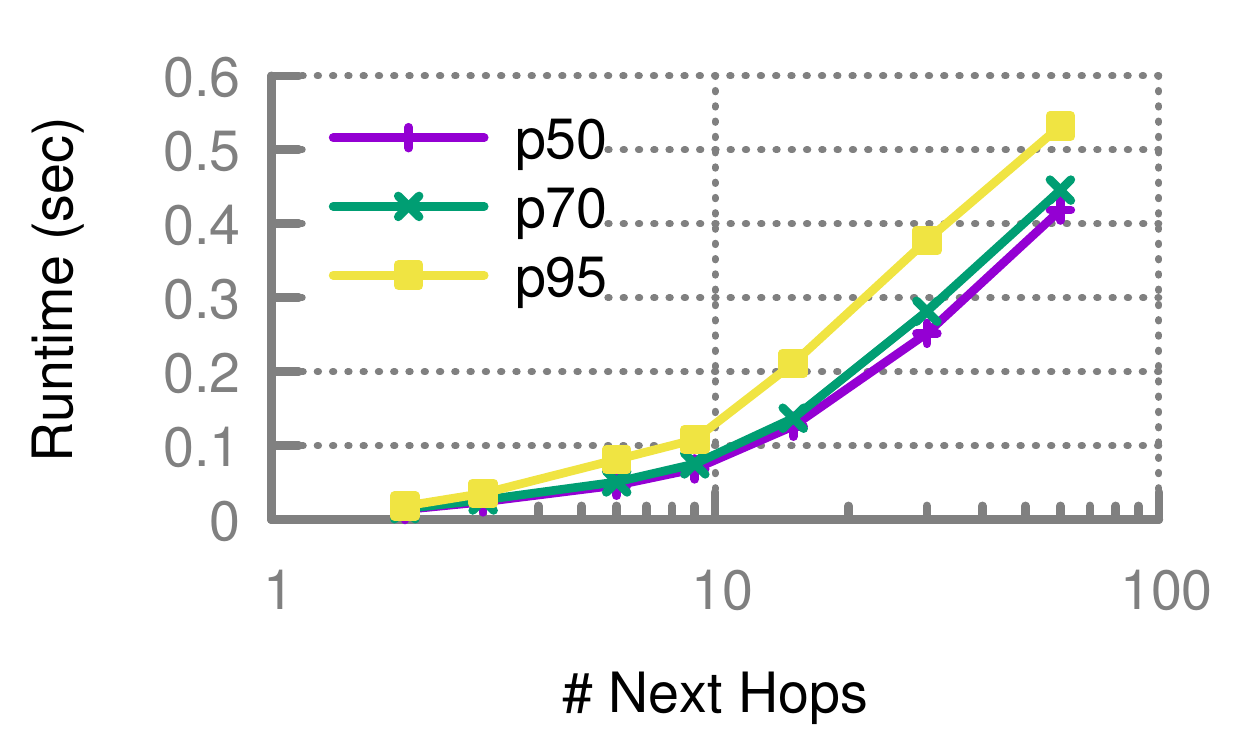}
  \vspace{-.2in}
  \label{fig:lp_nh}
 \end{subfigure}
  \begin{subfigure}[t]{.6\columnwidth}   \includegraphics[width=\textwidth]{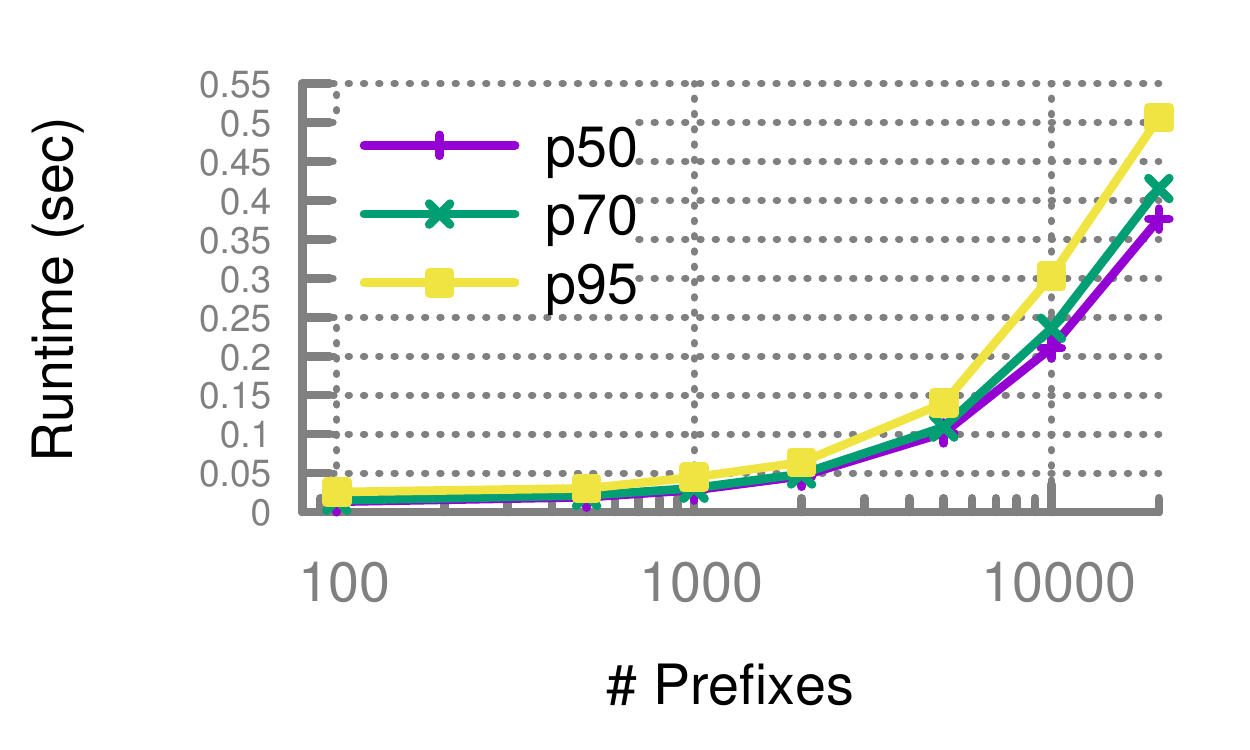}
  \label{fig:lp_dsts}
 \end{subfigure}
 \vspace{-.3in}
 \caption{\name is fast even when run with an increasing number of slots, next hops and destinations.}
\label{fig:run_timeN}
\end{figure*}

\myitem{Methodology:} We evaluate runtime, \ie the time the \solver takes to compute a forwarding state, across several scenarios with different numbers of prefixes, next-hops, and slots. For each scenario, we run >$5500$ experiments with four different objectives: performance, balance across next-hops, minimal number of steps, and all of these combined. We fix all but one of the three parameters (\ie prefixes, next-hops, and slots) to default values. By default, we set the number of prefixes to $800$ (corresponding to 80\% of the traffic of CAIDA.A); the number of next-hops to $3$, and the number of slots to be $200$ (corresponding to the minimum traffic-shift granularity of 0.5\% of the traffic per prefix). We report the median, $70$\textsuperscript{th}, and $95$\textsuperscript{th} percentile runtime as a function of each parameter in Fig.~\ref{fig:run_timeN}. We also group our experiments by objective and report median, $70$\textsuperscript{th}, and $95$\textsuperscript{th} percentile runtime in Table~\ref{tab:obj}.

\myitem{Key results:}
Fig~\ref{fig:run_timeN} shows that the $95$\textsuperscript{th}-percentile runtime is 0.25~sec for 22 slots per prefix (left), 0.1~sec for 10 next-hops per prefix (center), and 0.05~sec for 2K prefixes (right). As Table~\ref{tab:obj} shows, the runtime also depends on the complexity of the objective. The most efficient objective to solve for is minimizing the number of shifted slots, while the least efficient one, unsurprisingly, is the combination of all objectives. In nearly all cases, the \solver finishes in under one second.

%% file: case_study.tex
\section{Case studies}
\label{sec:case_study}
We validate \name's practicality and effectiveness in three steps. First, we prove that it is deployable by running it on a real testbed composed of Barefoot Tofino~\cite{tofino} switches. We then measure the benefits of running \name for 10 stub ASes. Finally, we highlight the effectiveness of \name in a larger testbed using P4\textsubscript{16}.  


\subsection{Hardware testbed}
We implement our hardware design (\S\ref{sec:tofino_implementation}) on a Barefoot Tofino Wedge 100BF-32X in which a control process pulls statistics every $1$ second, and updates routing accordingly.

\label{sec:data_eval}
\begin{figure*}[t]
 \centering
 \begin{subfigure}[t]{.3\textwidth}   \includegraphics[width=\textwidth]{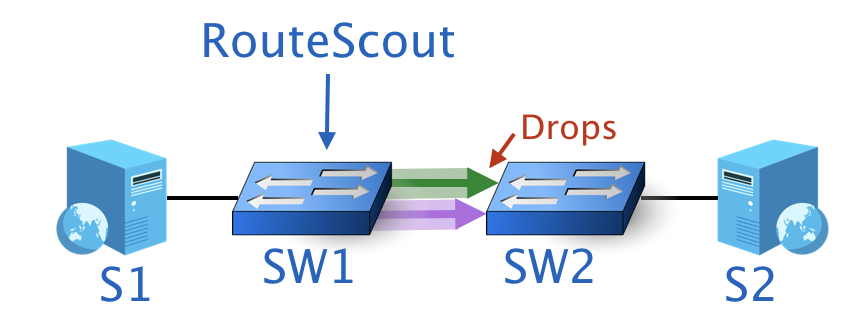}
  \caption{Two Tofinos set up two $s1$-$s2$ paths. $SW1$ runs \name and $SW2$ introduces loss in between the experiment.}
  \label{fig:testbed}
 \end{subfigure}
 \qquad
  \begin{subfigure}[t]{.3\textwidth}   \includegraphics[width=\textwidth]{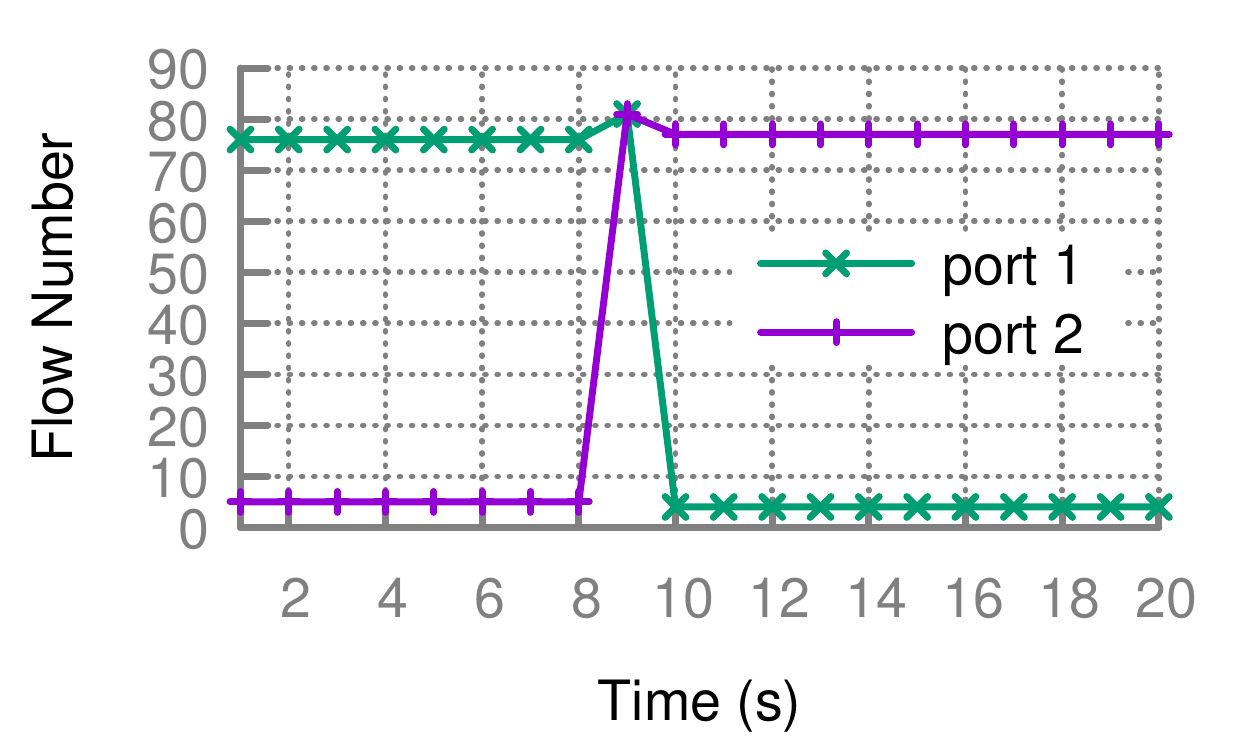}
  \caption{Number of flows routed via each alternative port changes after the increased loss is detected. Traffic shift takes <$2sec$. }
  \label{fig:flows}
 \end{subfigure}
  \qquad
  \begin{subfigure}[t]{.3\textwidth}   \includegraphics[width=\textwidth]{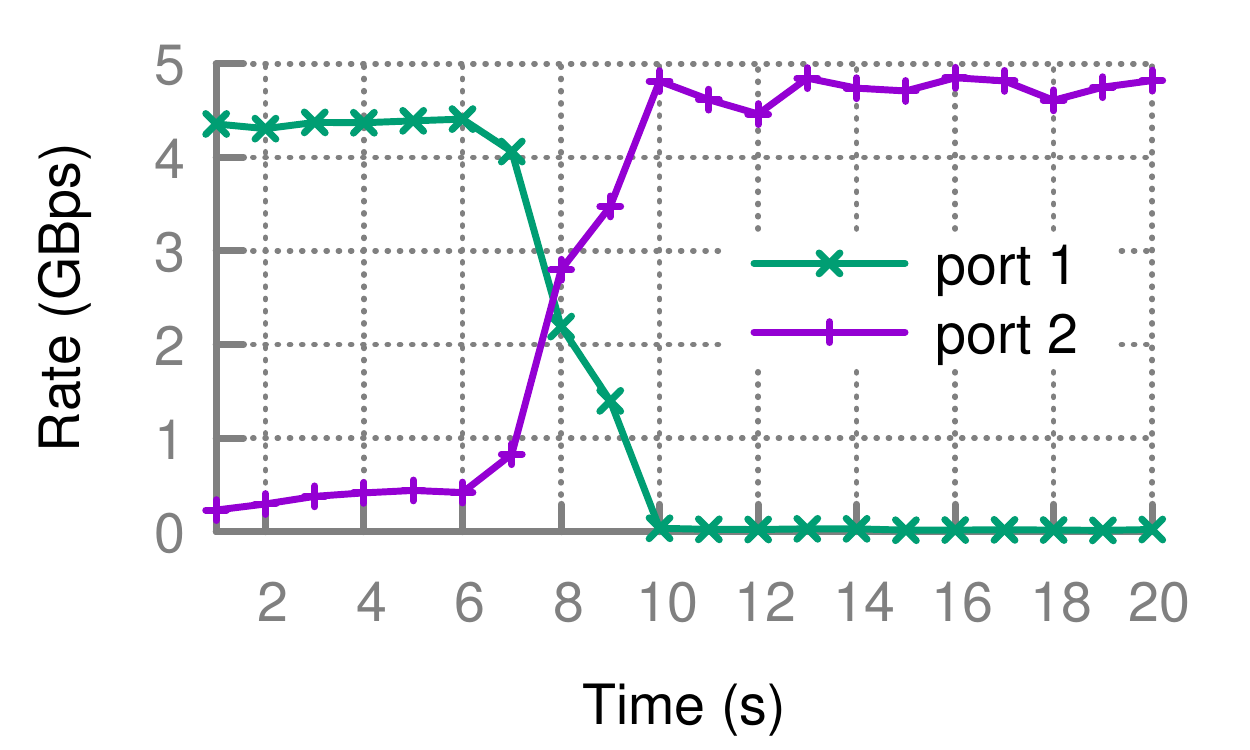}
  \caption{Bandwidth drop in port 1 is visible immediately after the loss is introduced and is clearer after \name reroutes traffic.}
  \label{fig:bits}
 \end{subfigure}
\vspace{-.1in}
\label{fig:run_time}
\caption{}
\end{figure*}

Our testbed (Fig.~\ref{fig:testbed}) has two Tofinos ($SW1$ and $SW2$) and two servers ($s1$ and $s2$). $SW1$--$SW2$ are connected to each other with two links via ports $1$ and $2$, creating two $s1$--$s2$ paths. $SW1$ runs \name and splits traffic to $s2$ across the two links. $SW2$ randomly drops a configurable portion of incoming packets matching on a specified ingress port.

We partition traffic to $s2$ into $16$ slots. Thus, the minimum portion of traffic \name can reroute/monitor is $1/16$ in this configuration.
(More generally, anything from $\frac{1}{2}-\frac{1}{2^{32}}$ is feasible.)
We assume the operator wants to minimize loss for traffic to $s2$. We also assume that the default next-hop for traffic to $s1$ is port 1, \ie the green (top) path. \name thus routes most traffic ($15/16$) on it, using one slot to probe the other path. 
We use $81$ iperf~\cite{iperf} client-servers pairs to generate $s1\rightarrow s2$ traffic. At time $t_1=7$~sec, we introduce $0.8\%$ loss on the top path using $SW2$.

Fig.~\ref{fig:flows} and Fig.~\ref{fig:bits} show how the flow-count and traffic at each port evolve. Initially, port 1 sees $76$ flows ($4.3$~Gbps) while port 2 sees only $5$ flows ($0.4$~Gbps). At $t_1$, loss starts, and bandwidth across the green path drops as TCP reacts. This is quickly detected ($<2sec$) by \name, which installs new rules to shift almost all the traffic to port $2$. \name could be made faster by (for instance) increasing the polling rate for statistics. A pure data-plane system that forgoes a controller will, of course, be even faster, but lose \name's flexibility in terms of optimization goals, and its stability.

\label{subsec:perfMeasure} 
\begin{figure}[t]
\centering
  \includegraphics[width=.85\columnwidth]{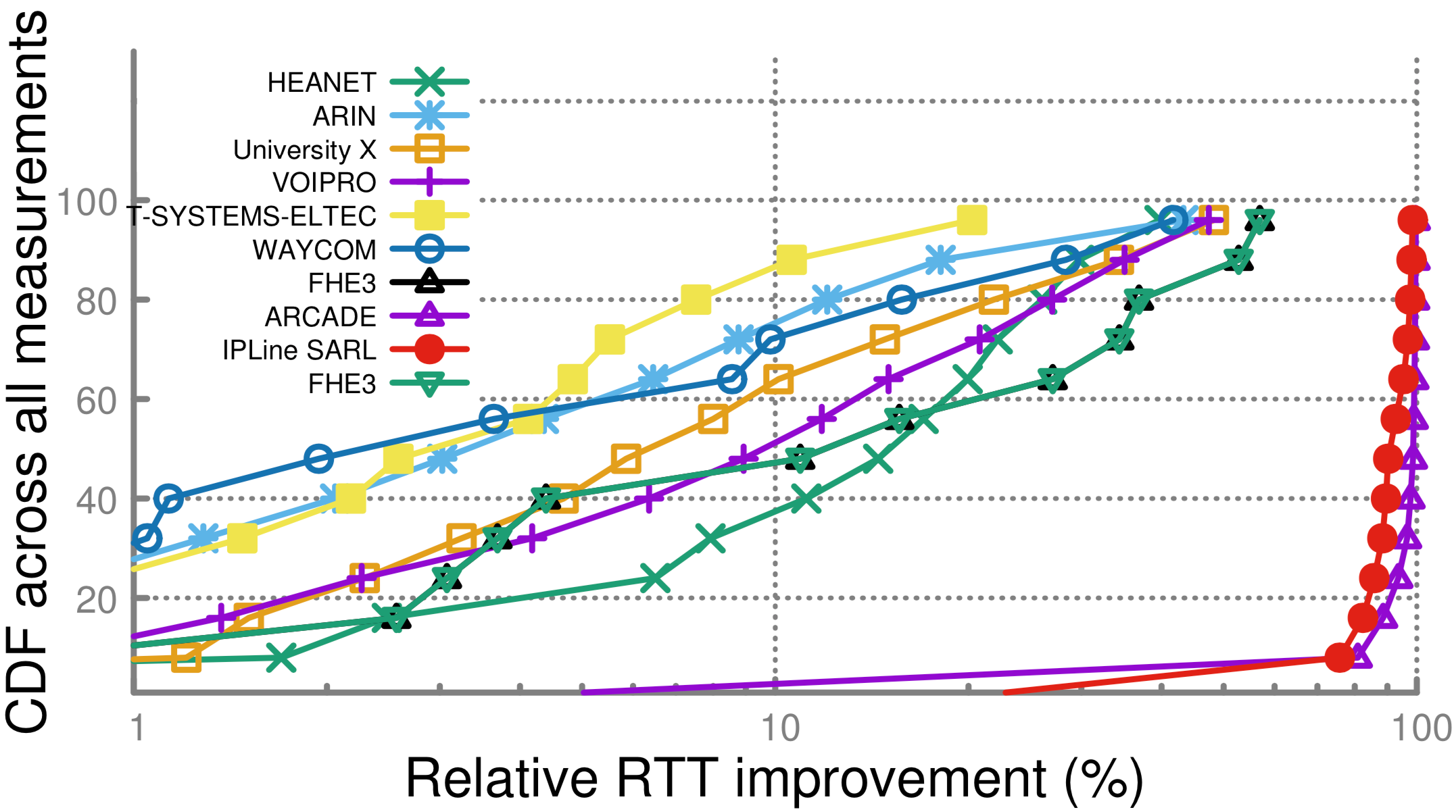}
\vspace{-.1in}
        \caption{CDF of the relative RTT improvement each source AS should expect from delay-aware routing.
        $8$ of the $10$ ASes could improve the latency of at least $20\%$ of the cases by $12$--$99\%$.} 
  \label{fig:rel_result}
\end{figure}

\subsection{Achievable gains in the wild}
\label{subsec:measurements}

Quantifying the gains provided by \name is challenging for three main reasons: \emph{(i)} one needs to control egress routing of the tested stub AS; \emph{(ii)} multiple stub ASes need to be tested for the results to be meaningful; \emph{(iii)} running the full system using previously collected traces is problematic as the traffic is not responsive to \name's operations (e.g. a lost packet will not be retransmitted).

To circumvent those limitations, we leverage \emph{(i)} the RIPE ATLAS
platform~\cite{ripe-atlas} which gives us access to multiple measurement probes
in many stub ASes all over the world; and \emph{(ii)} the fact that some stubs
host multiple probes whose traffic exits via different next-hops (due to hot
potato routing), and therefore take different paths.

In particular, we measure the delay difference among paths with same pair of source-AS and destination IP but different first nexthop. We believe this measurement is a reasonable proxy for the RTT improvement achievable with \name.
Every $~5$ minutes,\footnote{The maximum probing frequency allowed by RIPE ATLAS.} 
we perform $2$ concurrent traceroutes from $2$ probes in the same AS, to
each of the top-$50$ Alexa~\cite{alexa} destinations and report the diffrence in median delay observed by the two probes per pair of destination and 5-min interval iff they used a different next hop.
We perform this experiment for $24$ hours, and repeat it for 10 stub ASes.\footnote{The selection of ASes was done such that there is at least one pair of probes $a$, $b$ in AS$X$; which are geographically close to each other; and use different ASes, say $nextHopA$ and $nextHopB$ to reach the same destination prefix say $p$, which is among the 50 most popular Web destinations.} 
Fig.~\ref{fig:rel_result} shows the CDF of potential RTT improvement. Each line corresponds to a particular stub AS.

We find that 9/10 ASes could improve their RTTs in more than $35$\% of the cases by a $5$--$99$\% 
For $6$ ASes, RTT would improve by more than $21\%$ in at least $20$\% of the cases, while for $2$ of them, RTT improvement would exceed $97$\%.

\remove{
\begin{figure}[t]
\centering
  \includegraphics[width=\columnwidth]{figures/RipeAbsDiff.pdf}
\vspace{-.1in}
        \caption{CDF of the absolute RTT difference observed  when probing the same destinations from different vantage points.
        $6$ ASes could improve the latency of at least $20$\% of the cases by $23$--$246$ms.} 
  \label{fig:abs_result}
\end{figure}
}

\subsection{\name in a network}
We implement \name in the P4 behavioral model (\textsf{BMV2})~\cite{bmv2} using
$\sim$$900$ lines of P4\textsubscript{16}. We emulate a network scenario
with a stub that runs \name and $10$ destination networks towards each of which it has $3$ next-hops.
The network scenario has $14$ ASes, and $33$ $10$ Mbps AS-to-AS links.
The end-end delays are configured based on the latency differences observed in our RIPE experiments (\S\ref{subsec:measurements}). We assume that BGP has selected the first next hop for all prefixes. The goal of \name's operator is to minimize delay.

\begin{figure}[t]
\centering
  \includegraphics[width=.85\columnwidth]{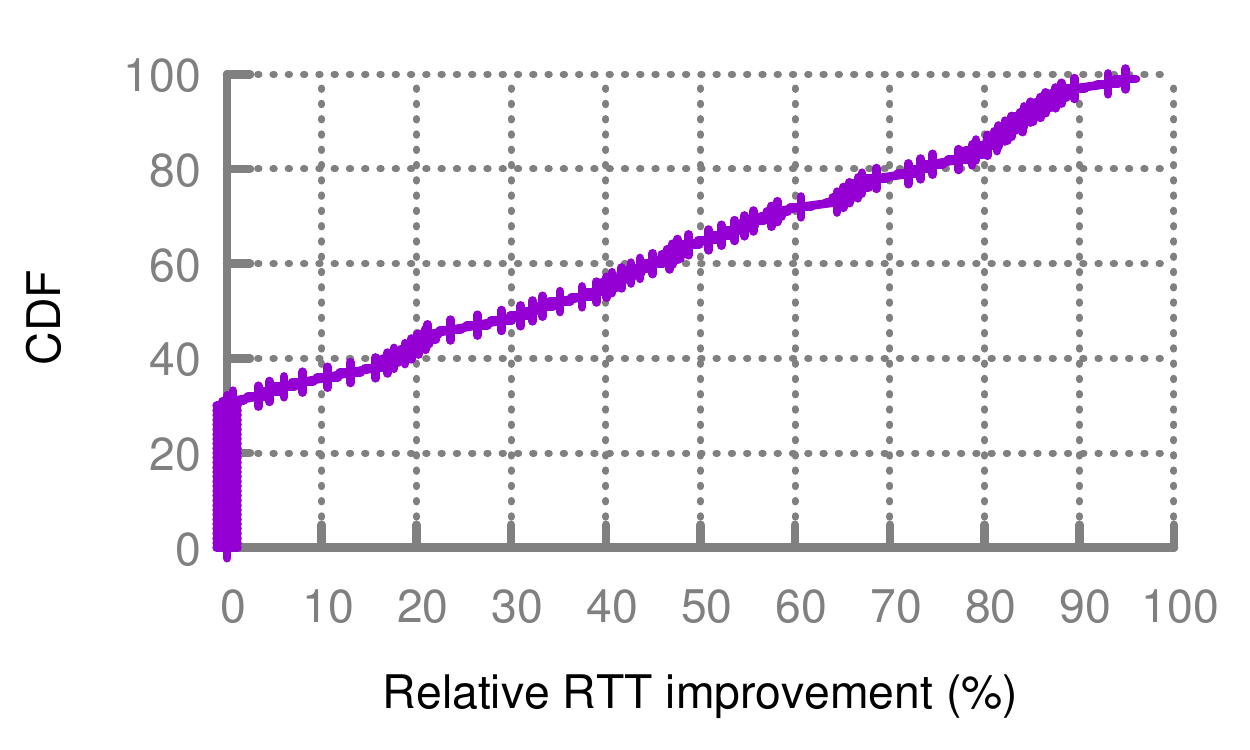}
\vspace{-.1in}
  \caption{CDF of \% delay improvement with \name.}
 \label{fig:minicdf}
\end{figure}

We use D-ITG~\cite{DBLP:journals/cn/BottaDP12} to create
$10$ TCP flows of constant rate to each of the destinations, resulting in $0.2$
Mbps of aggregated traffic. We configure \name to use $50$ slots in total; as
all prefixes drive the same traffic volume, each gets $5$ slots. We run the
experiment $10$ times and report (Fig.~\ref{fig:minicdf}) the CDF of
improvement on the average end-end delay compared with the initial state. We
see that \name improves the delay in half of the cases by $32\%$ or more.

%% file: conclusion.tex
\section{Conclusion}
\name is a modern answer to the old problem of performance-aware Internet
routing. Leveraging the capabilities of programmable switches, \name
continually and accurately monitors paths performance at scale with low
compute, memory, and bandwith footprints. Based on these measurements, \name
control plane then reroute traffic along policy-equivalent paths, fulfilling
the operators' objectives. \name is BGP-compatible, deployable without
coordination across ASes and without network-wide updates, improving Internet
routing one switch at a time.
